\documentclass[12pt]{article}
\usepackage{epsfig, epsf, graphicx}
\usepackage{pstricks, pst-node, psfrag}
\usepackage{amsmath,bm,amssymb}
\usepackage{verbatim,enumerate}
\usepackage{rotating, lscape,natbib}
\usepackage{setspace}
\usepackage{tikz}
\usetikzlibrary{arrows,calc,tikzmark}
\usepackage{hyperref}
\usepackage{float}
\usepackage{multirow}
\usepackage{algpseudocode}
\usepackage{algorithm}
\usepackage{array}
\usepackage{mathrsfs}
\usepackage{chngcntr}
\usepackage{cleveref}
\usepackage[caption=false]{subfig}
\usepackage[normalem]{ulem}
\usepackage{scrextend}

\counterwithin{algorithm}{subsection}

\def\boxit#1{\vbox{\hrule\hbox{\vrule\kern6pt
          \vbox{\kern6pt#1\kern6pt}\kern6pt\vrule}\hrule}}

\def\bse{\begin{eqnarray*}}
\def\ese{\end{eqnarray*}}
\def\ee{\end{eqnarray}}
\def\eq{\end{equation}}
\def\bse{\begin{eqnarray*}}
\def\ese{\end{eqnarray*}}

\newcommand{\bU}{\mathbf{U}}

\newcommand{\bV}{\mathbf{V}}

\newcommand{\bC}{\mathbf{C}}
\newcommand{\ba}{\mathbf{a}}
\newcommand{\bA}{\mathbf{A}}
\newcommand{\bb}{\mathbf{b}}
\newcommand{\bB}{\mathbf{B}}

\newcommand{\bI}{\mathbf{I}}

\newcommand{\bL}{\mathbf{L}}

\newcommand{\bmu}{\boldsymbol{\mu}}

\newcommand{\bxi}{\boldsymbol{\xi}}

\newcommand{\bsigma}{\boldsymbol{\sigma}}
\newcommand{\bSigma}{\mathbf{\Sigma}}

\newcommand{\bPhi}{\mathbf{\Phi}}

\newcommand{\btheta}{\boldsymbol{\theta}}

\newcommand{\bx}{\mathbf{x}}
\newcommand{\bX}{\mathbf{X}}
\newcommand{\by}{\mathbf{y}}
\newcommand{\bY}{\mathbf{Y}}
\newcommand{\bz}{\mathbf{z}}
\newcommand{\bZ}{\mathbf{Z}}
\newcommand{\bw}{\mathbf{w}}

\newcommand{\bh}{\mathbf{h}}

\newcommand{\0}{\mathbf{0}}

\newcommand*\diff{\mathop{}\!\mathrm{d}}

\graphicspath{{./Figures/}}
\begin{document}

\thispagestyle{empty} \baselineskip=28pt \vskip 5mm
\begin{center} {\Huge{\bf Exploiting Low Rank Covariance Structures for Computing High-Dimensional Normal and Student-$t$ Probabilities}}
	
\end{center}

\baselineskip=12pt \vskip 10mm

\begin{center}\large
Jian Cao \footnote[1]
{
\label{fn:kaust}
\baselineskip=10pt CEMSE Division, Extreme Computing Research Center,
King Abdullah University of Science and Technology,
Thuwal 23955-6900, Saudi Arabia.\\
E-mail: \{jian.cao, marc.genton, david.keyes\}@kaust.edu.sa\\
This research was supported by 
King Abdullah University of Science and Technology (KAUST).
}, Marc G.~Genton \footref{fn:kaust}, David E. Keyes \footref{fn:kaust},
and George M. Turkiyyah \footnote[2]{
\baselineskip=10pt Department of Computer Science, American University of Beirut, Beirut, Lebanon.\\
E-mail: gt02@aub.edu.lb\\}
\baselineskip=12pt \vskip 10mm
\today
\end{center}

\baselineskip=17pt \vskip 10mm

\begin{center}
{\large{\bf Abstract}}
\end{center}

We present a preconditioned Monte Carlo method for computing high-dimensional multivariate normal and Student-$t$ probabilities arising in spatial statistics. The approach combines a tile-low-rank representation of covariance matrices with a block-reordering scheme for efficient Quasi-Monte Carlo simulation. The tile-low-rank representation decomposes the high-dimensional problem into many diagonal-block-size problems and low-rank connections. The block-reordering scheme reorders between and within the diagonal blocks to reduce the impact of integration variables from right to left, thus improving the Monte Carlo convergence rate. Simulations up to dimension $65{,}536$ suggest that the new method can improve the run time by an order of magnitude compared with the hierarchical Quasi-Monte Carlo method and two orders of magnitude compared with the dense Quasi-Monte Carlo method. Our method also forms a strong substitute for the approximate conditioning methods as a more robust estimation with error guarantees. An application study to wind stochastic generators is provided to illustrate that the new computational method makes the maximum likelihood estimation feasible for high-dimensional skew-normal random fields. 

\baselineskip=14pt

\par\vfill\noindent

\clearpage\pagebreak\newpage \pagenumbering{arabic}
\baselineskip=26pt

\section{Introduction}
\label{sec:intro}

The multivariate normal (MVN) probability appears frequently in statistical applications. For example, the probability density functions of several skew-normal \citep{genton2004skew, azzalini2013skew, arellano2006unified} and Bayesian probit \citep{durante2019conjugate} models involve MVN cumulative distribution functions. It is also needed in computing the excursion and contour regions discussed in \cite{bolin2015excursion}. For many of these applications, the MVN probability is regarded as a bottleneck and approximations to the covariance matrix are often applied in high dimensions. The MVN probability is one example of numerical integration, in which the quadrature-based methods are typically not applicable in hundreds of dimensions. The Monte-Carlo-based methods are more flexible but their convergence rate is subject to several factors. In this paper, we aim to reduce the time costs and extend the limits for computing MVN probabilities.

The prevalent algorithm for computing MVN probabilities is based on the separation-of-variable (SOV) technique \citep{genz1992numerical}, which converts the integration region to the unit hypercube to improve the convergence rate. This method is more robust than its improved variants but has poor scalability, with the costs of $O(n^3)$ for the Cholesky factorization and $O(n^2)$ per MC sample, where $n$ is the MVN problem dimension. State-of-the-art methods for computing MVN probabilities include the hierarchical Quasi-Monte Carlo (QMC) method \citep{GKT2016hmvn}, the minimax tilting method \citep{botev2017normal}, the two-step method \citep{azzimonti2018estimating}, and the hierarchical conditioning method \citep{CGKT2018hcmvn}. The hierarchical QMC method reduced the costs per sample through the hierarchical representation \citep{hackbusch2015hierarchical} of the Cholesky factor. Its drawback is its incompatibility with variable reordering and hence, its inability to benefit from an improved convergence rate. The minimax tilting method significantly improves the convergence rate with importance sampling but needs to solve an expensive optimization with $O(n)$ parameters for the proposal density. The two-step method decomposes a high-dimensional MVN probability into a low-dimensional one and a high-dimensional residual, which is only applicable to orthant MVN probabilities that have constant upper and lower integration limits. \cite{CGKT2018hcmvn} used the hierarchical representation and the conditioning technique that samples the integrand only once to achieve high computation efficiency. However, this hierarchical conditioning method only provides crude probability estimates without any error estimation.

This paper builds on the original SOV method in \cite{genz1992numerical} and introduces a variant that has better performance than the hierarchical QMC method in \cite{GKT2016hmvn}. Specifically, we combine the SOV method with the tile-low-rank (TLR) representation \citep{weisbecker2013improving, mary2017block, akbudak2017tile}, which improves efficiency from two aspects. First, the TLR representation is compatible with block-wise variable reordering and hence, benefits from a higher convergence rate. Secondly, the memory footprint of the Cholesky factor under the TLR representation can be smaller than that under the hierarchical representation, indicating lower costs per QMC sample. In this paper, we only compare our methods with the hierarchical QMC method in \cite{GKT2016hmvn} because the other three state-of-the-art methods are not directly based on the SOV algorithm. As another extension, we propose an iterative version of the original block reordering in \cite{CGKT2018hcmvn} that further improves the convergence rate and performs the Cholesky factorization simultaneously. The corresponding algorithm for multivariate Student-$t$ (MVT) probabilities is also developed. Finally, we demonstrate the capability of our methods in tens of thousands of dimensions with two maximum likelihood estimation (MLE) studies based on simulated data and a wind dataset.

The remainder of this paper is structured as follows. In \Cref{sec:dense_lr_qmc}, we review the SOV algorithm \citep{genz2009computation} for MVN and MVT problems and describe the dense QMC algorithms for both probabilities. In \Cref{sec:blk_reorder}, we show that the TLR representation is more aligned with block-wise variable reordering than hierarchical representations. Additionally, an improved version of the block reordering from \cite{CGKT2018hcmvn} is proposed. In \Cref{sec:simulation}, we compare the dense QMC method, the hierarchical QMC method, and the TLR QMC methods with a focus on high-dimensional MVN and MVT probabilities. In \Cref{sec:application}, we estimate the parameters for simulated high-dimensional skew-normal random fields as well as fit the skew-normal model to a large wind speed dataset of Saudi Arabia to demonstrate the usage of our methods. Finally, \Cref{sec:conclusion_tlrmvnmvt} concludes the paper. The execution times in this paper are measured on a 4-core Intel Core i7 CPU with 64 GB memory without parallelization.

\section{SOV for MVN and MVT Probabilities}
\label{sec:dense_lr_qmc}

The SOV technique transforms the integration region into the unit hypercube, where efficient QMC rules can improve the convergence rate. The SOV of MVN probabilities is based on the Cholesky factor of the covariance matrix \citep{genz1992numerical} and this naturally leads to the second form of SOV for MVT probabilities \citep{genz2002comparison}. The two forms of SOV for MVT probabilities have been derived in \cite{genz1992numerical} and \cite{genz2002comparison}. In this paper, we summarize the derivations for completeness.
\subsection{SOV for MVN integrations}
\label{subsec:sov_mvn}

We denote an $n$-dimensional MVN probability with $\Phi_n(\ba,\bb;\bmu,\bSigma)$, where $(\ba,\bb)$ defines a hyperrectangle-shaped integration region, $\bmu$ is the mean vector, and $\bSigma$ is the covariance matrix. The MVN probability has the form:
\begin{equation}
 \Phi_n(\ba,\bb;\bmu,\bSigma) = 
 \int_{\mathbf{a}-\bmu}^{\mathbf{b}-\bmu}{\frac{1}{\sqrt{(2\pi)^n |\mathbf{\Sigma}|}}\exp \left(-\frac{1}{2}\mathbf{x}^{\top}\mathbf{\Sigma}^{-1}\mathbf{x} \right)}\diff \mathbf{x}.
\label{eq:mvn}	
\end{equation}
Without loss of generality, we set $\bmu = \mathbf{0}$ and denote the $n$-dimensional MVN probability with $\bPhi_n(\ba,\bb;\bSigma)$. We use $\bL$ to represent the lower Cholesky factor of $\bSigma = \bL \bL^{\top}$ and $l_{ij}$ to represent the element on the $i$-th row and $j$-th column of $\bL$. Following the procedure in \cite{genz1992numerical}, we can transform $\Phi_n(\ba,\bb;\bSigma)$ into:
\begin{equation}
 \label{equ:sov_mvn}
 \Phi_n(\ba,\bb;\bSigma) = 
 (e_1 - d_1) \int_0^1 (e_2 - d_2) \cdots \int_0^1 (e_n - d_n) \int_0^1 \diff \bw,
\end{equation}
where $d_i = \Phi\{(a_i - \sum_{j=1}^{i-1}l_{ij}y_j)/l_{ii}\}$, $e_i = \Phi\{(b_i - \sum_{j=1}^{i-1}l_{ij}y_j)/l_{ii}\}$, $y_j = \Phi^{-1}\{d_j + w_j(e_j - d_j)\}$, and $\Phi(\cdot)$ is the cumulative distribution function (CDF) of the standard normal distribution.

The integration region is transformed into $[0,1]^n$ and efficient sampling rules can be applied to simulate $\bw$, although the integrand is difficult to compute in parallel because $d_i$ and $e_i$ depend on $\{y_j, j = 1, \ldots, i-1\}$ while $y_i$ depends on $d_i$ and $e_i$. Only univariate standard normal probabilities and quantile functions are needed, which can be readily obtained with the high efficiency of scientific computing libraries, for example, the Intel MKL. The Cholesky factorization has a complexity of $O(n^3)$ but modern CPUs and libraries have been developed to handle matrices with more than $10{,}000$ dimensions with ease.

We use `mvn' to denote the integrand function of \Cref{equ:sov_mvn}, whose pseudocode was originally proposed in \cite{genz1992numerical}. Because the `mvn' function is also the subroutine in other functions of this paper, we summarize it here in \Cref{alg:mvn}. The algorithm returns $P$, the probability estimate from one sample and $\by$ whose coefficients are described in \Cref{equ:sov_mvn}.
\begin{algorithm}
\caption{QMC for MVN probabilities}
\label{alg:mvn}
\begin{algorithmic}[1]
\State{mvn($\bL, \ba, \bb, \bw$)}
\State{$n \gets \mbox{dim}(\bL)$, $s \gets 0$, $\by \gets \mathbf{0}$, and $P \gets 1$}
\For{$i = 1:n$} 
\If{$i > 1$}
\State{$s \gets \bL(i,1:i-1) \by(1:i-1)$}
\EndIf
\State{$a' \gets \frac{a_i - s}{C_{i,i}}$, and $b' \gets \frac{b_i - s}{C_{i,i}}$}
\State{$y_i \gets \Phi^{-1}[w_{i}\{\Phi(b') - \Phi(a')\}]$}
\State{$P \gets P \cdot \{\Phi(b')-\Phi(a')\}$}
\EndFor
\State \Return $P$ and $\by$
\end{algorithmic}
\end{algorithm}
\noindent
Keeping $\ba$, $\bb$, and $\bL$ unchanged, the mean and standard deviation of the outputs $P$ from a set of well designed $\bw$, usually conforming to a Quasi-Monte Carlo rule, form the probability and error estimates. In our implementation, we employ the Richtmyer Quasi-Monte Carlo rule \citep{richtmyer1951evaluation}, where the batch number is usually much smaller than the batch size. 

\subsection{SOV for MVT integrations}
\label{subsec:sov_mvt}

We denote an $n$-dimensional MVT probability with $T_n(\ba,\bb;\bmu,\bSigma,\nu)$, where $\nu$ is the degrees of freedom. Here, $\bmu$ is the mean vector and $\bSigma$ is the scale matrix. To simplify the notations, $\bmu$ is again assumed to be $\0$. There are two common equivalent definitions for $T_n$, of which the first one is:
\begin{equation}
 \label{equ:mvt_def_form1}
 T_n(\ba,\bb;\bSigma,\nu) = \frac{\Gamma(\frac{\nu+n}{2})}{\Gamma(\frac{\nu}{2})\sqrt{| \bSigma | (\nu \pi)^n}}
 \int_{a_1}^{b_1}\cdots\int_{a_n}^{b_n}\left( 1 + \frac{\bx^{\top} \bSigma^{-1} \bx}{\nu} \right)^{-\frac{\nu+n}{2}} \diff \bx,
\end{equation}
where $\Gamma(\cdot)$ is the gamma function. Based on this definition, \cite{genz1999numerical} transformed the integration into the $n$-dimensional hypercube, where the inner integration limits depend on the outer integration variables. However, the integration needs to compute the CDF and the quantile function of the univariate Student-$t$ distribution at each integration variable. A second equivalent form defines $T_n$ as a scale mixture of the MVN probability, specifically:
\begin{subequations}
\begin{align}
 \label{equ:mvt_def_form2}
 T_n(\ba,\bb;\bSigma,\nu) &= 
 \frac{2^{1-\frac{\nu}{2}}}{\Gamma(\frac{\nu}{2})} \int_{0}^{\infty}s^{\nu-1}e^{-s^2/2}\Phi_n \left( \frac{s\ba}{\sqrt{\nu}}, \frac{s\bb}{\sqrt{\nu}}; \bSigma \right) \diff s, \\
 \label{equ:mvt_def_form2_trf}
 &= 
 E\left[\Phi_n \left( \frac{S\ba}{\sqrt{\nu}}, \frac{S\bb}{\sqrt{\nu}}; \bSigma \right)\right].
\end{align}
\end{subequations}
The density of a $\chi$-distribution random variable, $S$, with degrees of freedom $\nu$, is exactly $\frac{2^{1-\frac{\nu}{2}}}{\Gamma(\frac{\nu}{2})} s^{\nu-1}e^{-s^2/2}$, $s>0$. Thus, $T_n(\ba,\bb;\bSigma,\nu)$ can be also written as \Cref{equ:mvt_def_form2_trf}. The integrand boils down to the MVN probability discussed in the previous section. Hence, we can apply a Quasi-Monte Carlo rule in the $(n+1)$-dimensional hypercube to approximate this expectation, where only the CDF and the quantile function of the univariate standard normal distribution are involved. It is worth pointing out that considering $T_n$ as a one-dimensional integration of $\Phi_n$ and applying quadrature is much more expensive than integrating directly in $(n+1)$ dimensions.

We describe the integrand functions based on the two SOV schemes in \Cref{alg:mvt1} and \Cref{alg:mvt2}, corresponding to \Cref{equ:mvt_def_form1} and \Cref{equ:mvt_def_form2}, respectively.
\begin{algorithm}
\caption{QMC for MVT probabilities based on \Cref{equ:mvt_def_form1}}
\label{alg:mvt1}
\begin{algorithmic}[1]
\State{mvt\_sov($\bL, \ba, \bb, \nu, \bw$)}
\State $n \gets \mbox{dim}(\bL)$, $s \gets 0$, $ssq \gets 0$, $\by \gets \mathbf{0}$, and $P \gets 1$
\For{$i = 1:n$} 
\If{$i > 1$}
\State $s \gets \bL(i,1:i-1) \by(1:i-1)$
\EndIf
\State $a' \gets \frac{a_i-s}{\bL_{i,i} \cdot \sqrt{\nu+ssq} \cdot (\nu+i)}$ and $b' \gets \frac{b_i-s}{\bL_{i,i} \cdot \sqrt{\nu+ssq} \cdot (\nu+i)}$
\State $y_i \gets T_{\nu+i}^{-1} \left[ w_i\left\{T_{\nu+i}(b')-T_{\nu+i}(a')\right\} + T_{\nu+i}(a') \right] \cdot \sqrt{\frac{\nu + ssq}{\nu + i}}$
\State $P \gets P \cdot \left\{T_{\nu+i}(b')-T_{\nu+i}(a')\right\}$
\State $ssq \gets ssq + y_i^2$
\EndFor
\State \Return $P$
\end{algorithmic}
\end{algorithm}
\begin{algorithm}
\caption{QMC for MVT probabilities based on \Cref{equ:mvt_def_form2}}
\label{alg:mvt2}
\begin{algorithmic}[1]
\State{mvt\_scale($\bL, \ba, \bb, \nu, w_0 , \bw$)}
\State $\ba' \gets \frac{\chi^{-1}_\nu(w_0)}{\sqrt{\nu}} \ba$, $\bb' \gets \frac{\chi^{-1}_\nu(w_0)}{\sqrt{\nu}} \bb$
\State \Return $\mbox{mvn}(\bL,\ba',\bb',\bw)$
\end{algorithmic}
\end{algorithm}
\noindent
\Cref{alg:mvt1} calls the univariate Student-$t$ CDF and the quantile function with an increasing value of degrees of freedom at each iteration whereas \Cref{alg:mvt2} relies on $(w_0,\bw)$ from an $(n+1)$-dimensional Quasi-Monte Carlo rule and calls the `mvn' kernel from \Cref{alg:mvn} with the scaled integration limits. We use single-quoted `mvn' and `mvt' to denote the corresponding algorithms to distinguish them from the uppercase MVN and MVT used for multivariate normal and Student-$t$ in this paper.
\begin{table}[t!]
{\footnotesize
\caption{Relative error and time of the three algorithms. `mvt\_sov', `mvt\_scale', and `mvn' refer to \Cref{alg:mvt1}, \Cref{alg:mvt2}, and \Cref{alg:mvn}. The upper row is the average relative estimation error and the lower row is the average computation time over $20$ replicates. The covariance matrix is generated from a 2D exponential kernel, $\exp(- \|\bh\| / \beta)$, where $\bh$ is the distance vector, based on $n$ locations on a perturbed grid in the unit square. The lower integration limits are $-\bm{\infty}$ and the upper limits are independently generated from $N(5.5, 1.25^2)$. The degrees of freedom $\nu$ for MVT probabilities are $10$. The Monte Carlo sample size is $10^{4}$.}
\label{tbl:mvt_cmp}
\begin{center}
\begin{tabular}{crrrrr}
\hline \noalign{\smallskip}
$n$ & $16$ & $64$ & $256$ & $1{,}024$ & $4{,}096$ \\
\noalign{\smallskip}\hline\noalign{\smallskip}
mvt\_sov & 
\begin{tabular}{@{}r@{}} $0.0\%$  \\ \textit{0.7s}\end{tabular}& 
\begin{tabular}{@{}r@{}} $0.2\%$  \\ \textit{3.0s}\end{tabular}& 
\begin{tabular}{@{}r@{}} $0.7\%$  \\ \textit{13.3s}\end{tabular}& 
\begin{tabular}{@{}r@{}} $1.4\%$  \\ \textit{58.7s}\end{tabular}& 
\begin{tabular}{@{}r@{}} $4.2\%$  \\ \textit{283.1s}\end{tabular}\\ 
mvt\_scale & 
\begin{tabular}{@{}r@{}} $0.0\%$  \\ \textit{0.0s}\end{tabular}& 
\begin{tabular}{@{}r@{}} $0.0\%$  \\ \textit{0.0s}\end{tabular}& 
\begin{tabular}{@{}r@{}} $0.2\%$  \\ \textit{0.2s}\end{tabular}& 
\begin{tabular}{@{}r@{}} $0.4\%$  \\ \textit{2.0s}\end{tabular}& 
\begin{tabular}{@{}r@{}} $1.3\%$  \\ \textit{40.8s}\end{tabular}\\ 
mvn & 
\begin{tabular}{@{}r@{}} $0.0\%$  \\ \textit{0.0s}\end{tabular}& 
\begin{tabular}{@{}r@{}} $0.0\%$  \\ \textit{0.0s}\end{tabular}& 
\begin{tabular}{@{}r@{}} $0.1\%$  \\ \textit{0.2s}\end{tabular}& 
\begin{tabular}{@{}r@{}} $0.4\%$  \\ \textit{2.0s}\end{tabular}& 
\begin{tabular}{@{}r@{}} $1.2\%$  \\ \textit{40.1s}\end{tabular}\\
\end{tabular} 
\end{center}
}
\end{table}

A numerical comparison between \Cref{alg:mvt1} and \Cref{alg:mvt2} is shown in \Cref{tbl:mvt_cmp}. The counterpart for MVN probabilities (\Cref{alg:mvn}) is included as a benchmark. The table indicates that the first definition as in \Cref{equ:mvt_def_form1} leads to an implementation slower by one order of magnitude. Additionally, the convergence rate from \Cref{equ:mvt_def_form1} is also worse than that from \Cref{equ:mvt_def_form2}. Although the univariate Student-$t$ CDF and quantile function are computed the same number of times as their standard normal counterparts, their computation takes much more time, likely because of the lack of optimized libraries, and produces lower accuracy. Due to its performance advantage, we refer to \Cref{alg:mvt2} as the `mvt' algorithm from this point on. It has negligible marginal complexity over the `mvn' algorithm since the only additional step is scaling the integration limits.

\section{Low-rank Representation and Reordering for MVN and MVT Probabilities}
\label{sec:blk_reorder}

\subsection{Overview}
\label{subsec:br_lr_overview}

\begin{figure}
	\hfill
	\subfloat{\includegraphics[width = 0.3\linewidth]{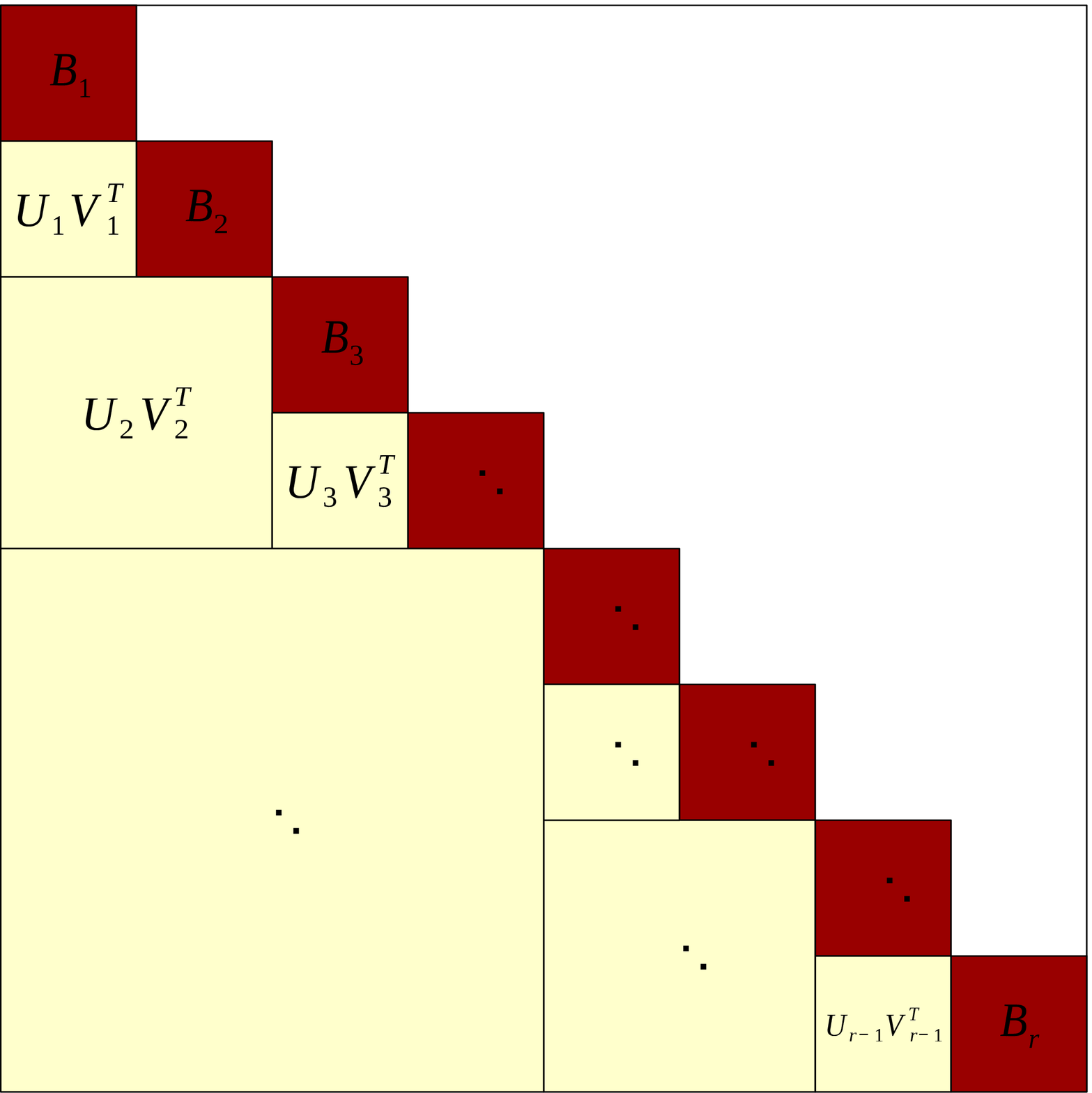}} 
	\hspace*{0.75in}
	\subfloat{\includegraphics[width = 0.3\linewidth]{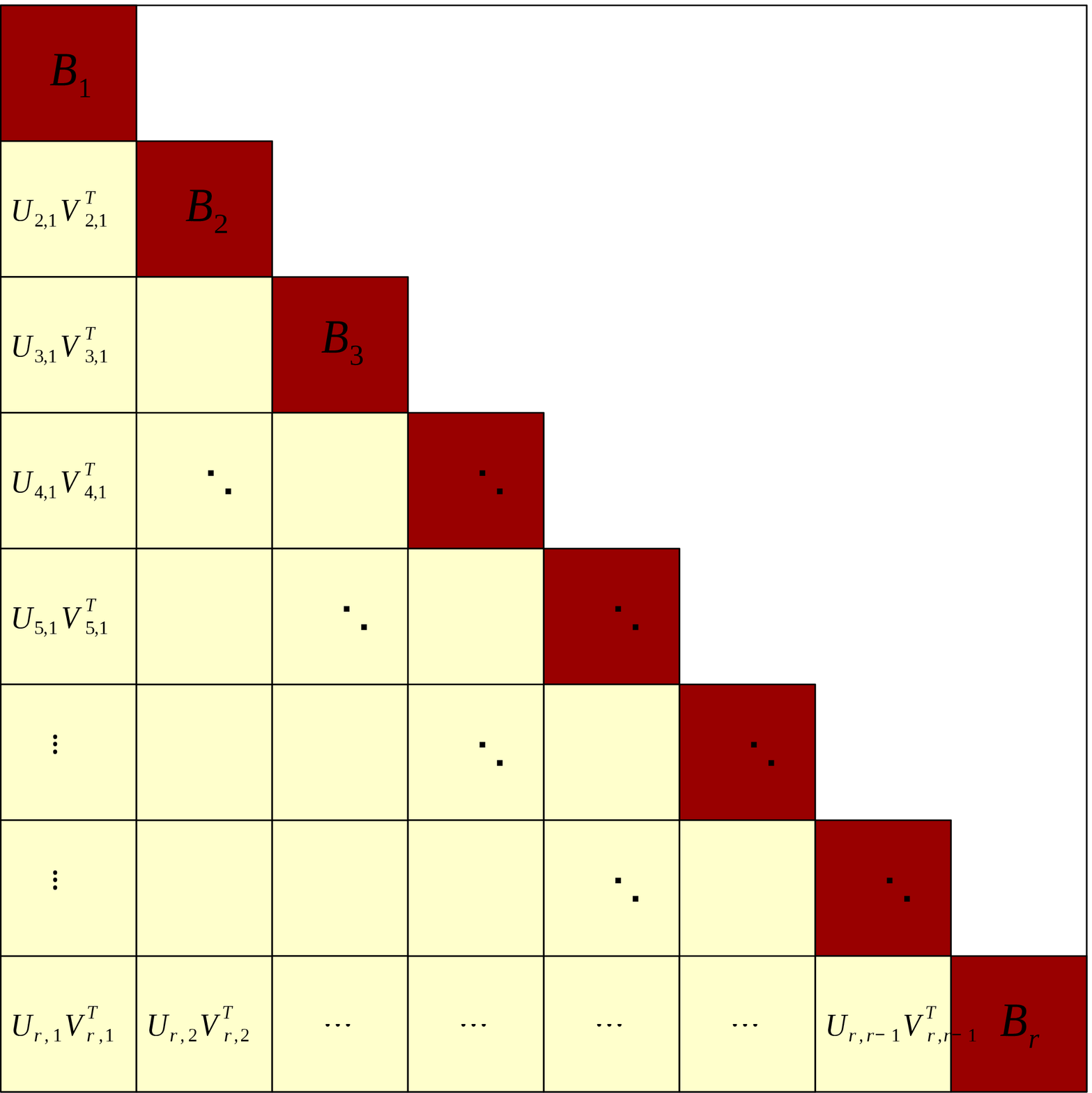}} 
	\hfill
	\caption{Structures of hierarchical (left) and tile-low-rank (right) matrices.}
	\label{fig:structure}
\end{figure}

More flexible than quadrature methods, Monte Carlo (MC) procedures provide several viable options for computing MVN and MVT probabilities. The cost of these computations depends on the product of the number of MC samples, $N$, needed to achieve a desired accuracy and the cost per MC sample. Under the standard dense representation of covariance, the computational complexity for each sample is $O(n^2)$ as shown in \Cref{alg:mvn} and \Cref{alg:mvt2}. \cite{GKT2016hmvn} proposed using the hierarchical representation for the Cholesky factor, illustrated in \Cref{fig:structure}, which reduced the complexity per sample to $O(kn \log n)$, where $k$ is a nominal local rank of the matrix blocks. Using nested bases in the hierarchical representation \citep{boukaram18b}, it is possible to reduce this cost further to an asymptotically optimal $O(kn)$. 

In this paper, we assume the covariance matrix is generated from a set of spatial locations and a covariance kernel. Small local ranks in off-diagonal blocks are obtained when the row cluster and the column cluster are well separated spatially, growing only weakly with the problem dimension, $n$. When the geometry is a subset of $\mathbb{R}^2$ or $\mathbb{R}^3$, a space-filling curve, or a spatial partitioning method in combination with a space-filling curve, may be used for indexing to keep the index distances reasonably consistent with the spatial distances. The spatial locations and the corresponding variables are then further divided into blocks (clusters) according to these indices to build the hierarchical representation. We also use the terms `cluster' to align with the literature of hierarchical matrices; see \cite{hackbusch2015hierarchical} for more details.

The optimal ordering for reducing the cost per Monte Carlo sample however is unfortunately generally not the optimal ordering for reducing the total number of samples $N$. A proper reordering scheme that takes into account the widths of the integration limits of the MVN and MVT probabilities can have a substantial effect on reducing the variance of the estimates, making the numerical methods far more effective relative to a default ordering \citep{schervish1984algorithm, genz2009computation}. \cite{trinh2015bivariate} analyzed ordering heuristics and found that a univariate reordering scheme, that sorts the variables so that the outermost integration variables have the smallest expected values, significantly increased the estimation accuracy. 
This heuristic was more effective overall than more expensive bivariate reordering schemes that might further reduce the number of samples needed. In \cite{CGKT2018hcmvn}, a block-reordering scheme was proposed under the hierarchical matrix representations used in high dimensions. Specifically, within each diagonal block $\bB_i$, univariate reordering was applied and the blocks were reordered based on their estimated probabilities using this univariate reordering scheme. This has less impact on the local ranks of the hierarchical structure than the reordering schemes discussed in \cite{trinh2015bivariate}.

The important point here is that these reordering schemes shuffle the variables based on their integration limits to achieve better convergence for the integration, measured by the number of samples needed to achieve the desired accuracy. They produce different orders from the geometry-oriented ordering obtained by spatial partitioning methods or space-filling curves. The reordering increases the local ranks $k$ of the hierarchical representation or broader low-rank representations, making the per-sample computation more expensive. 

In this paper, we seek a better middle ground between the geometry-oriented and the integration-oriented orderings by combining a block-reordering scheme with the TLR representation of covariance illustrated in \Cref{fig:structure}. We also introduce the TLR versions of the QMC algorithms for computing MVN and MVT probabilities.

\subsection{TLR as a practical representation for MVN and MVT}

\begin{figure}
  \centering
  \subfloat[HODLR \& Geom]{\label{fig:n1}{\includegraphics[width=0.29\textwidth]{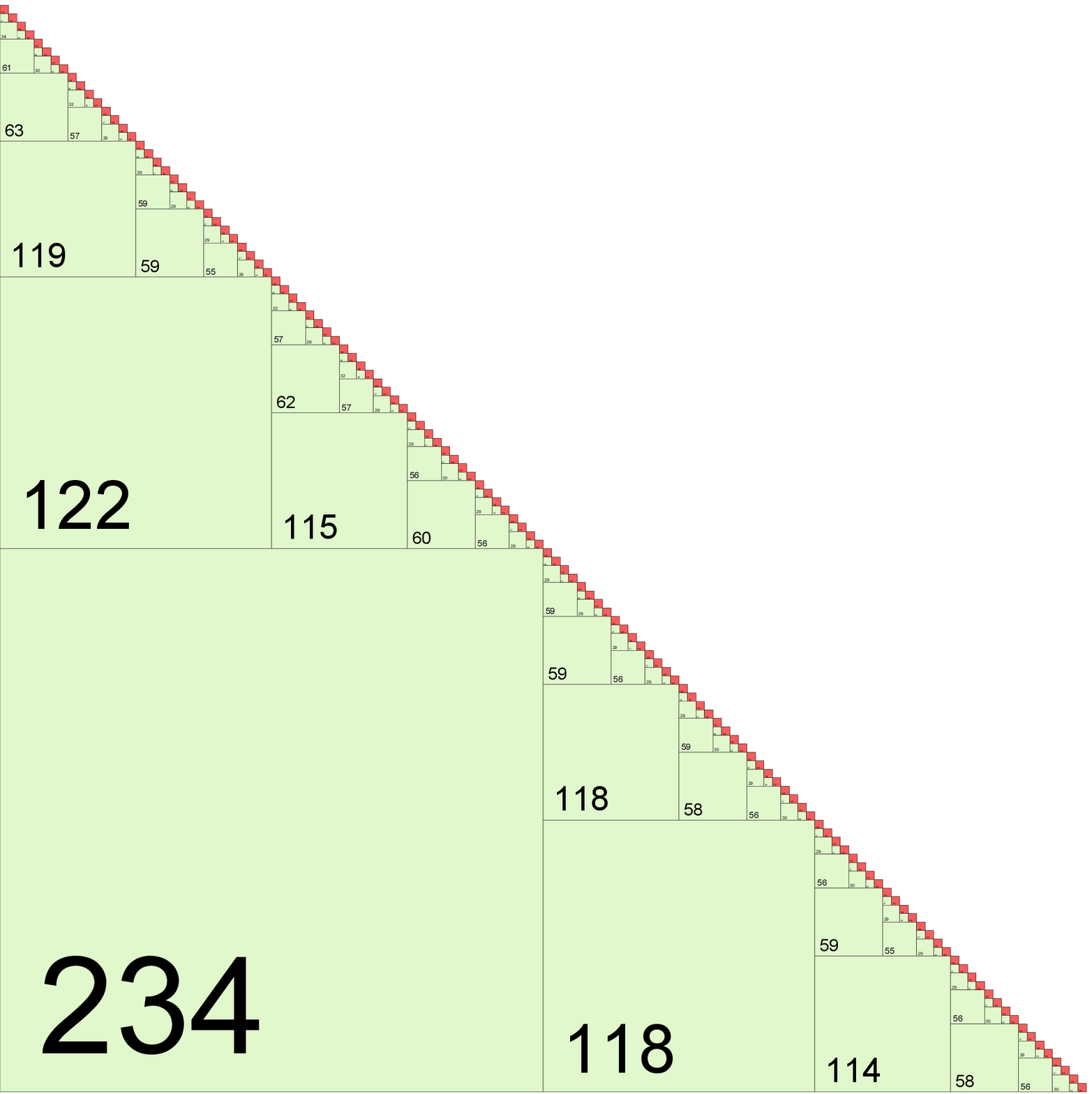}}}
  \hspace*{\fill}
  \subfloat[Standard \& Geom]{\label{fig:n2}{\includegraphics[width=0.29\textwidth]{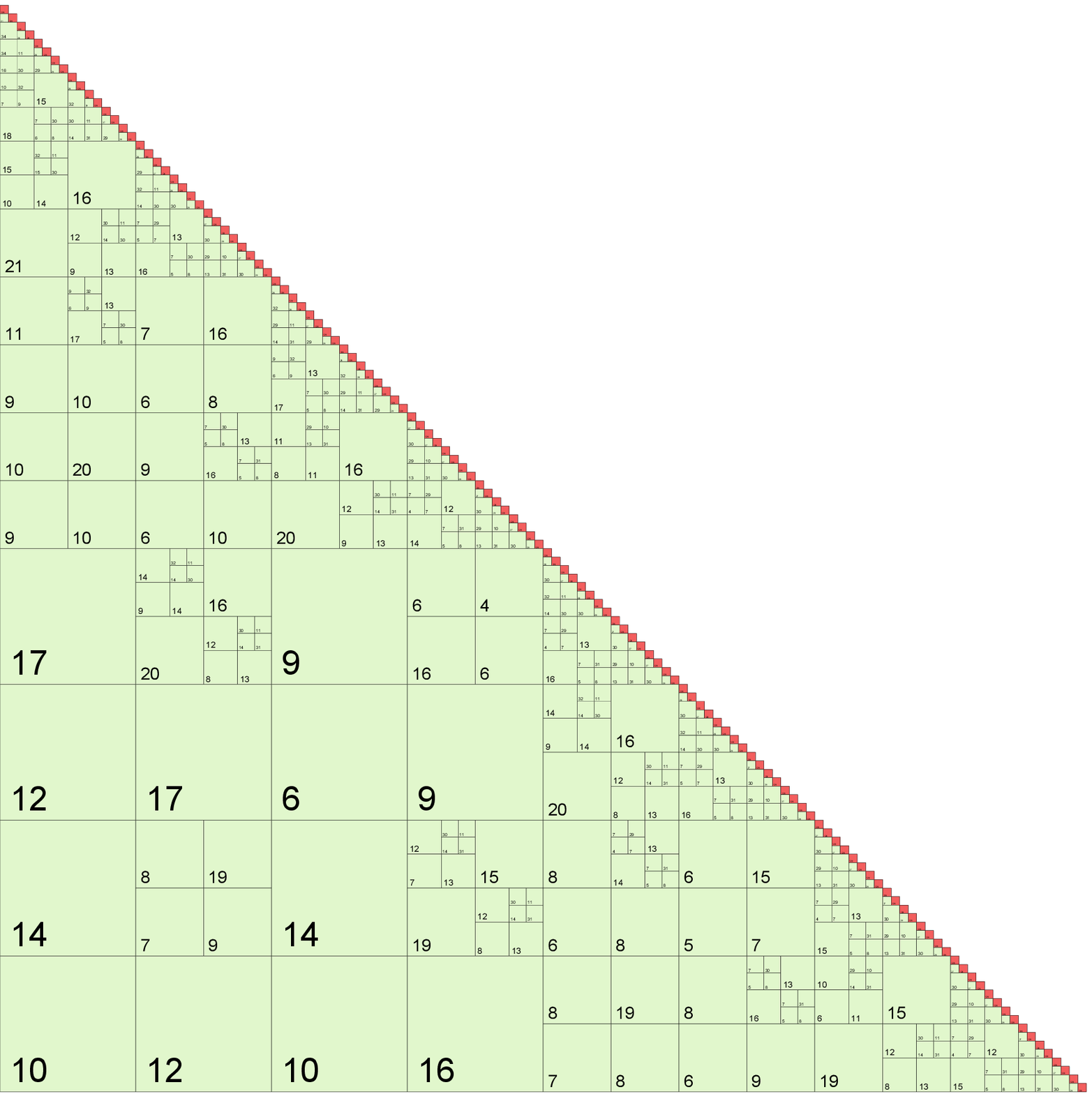}}}
  \hspace*{\fill}
  \subfloat[TLR \& Geom]{\label{fig:n2}{\includegraphics[width=0.29\textwidth]{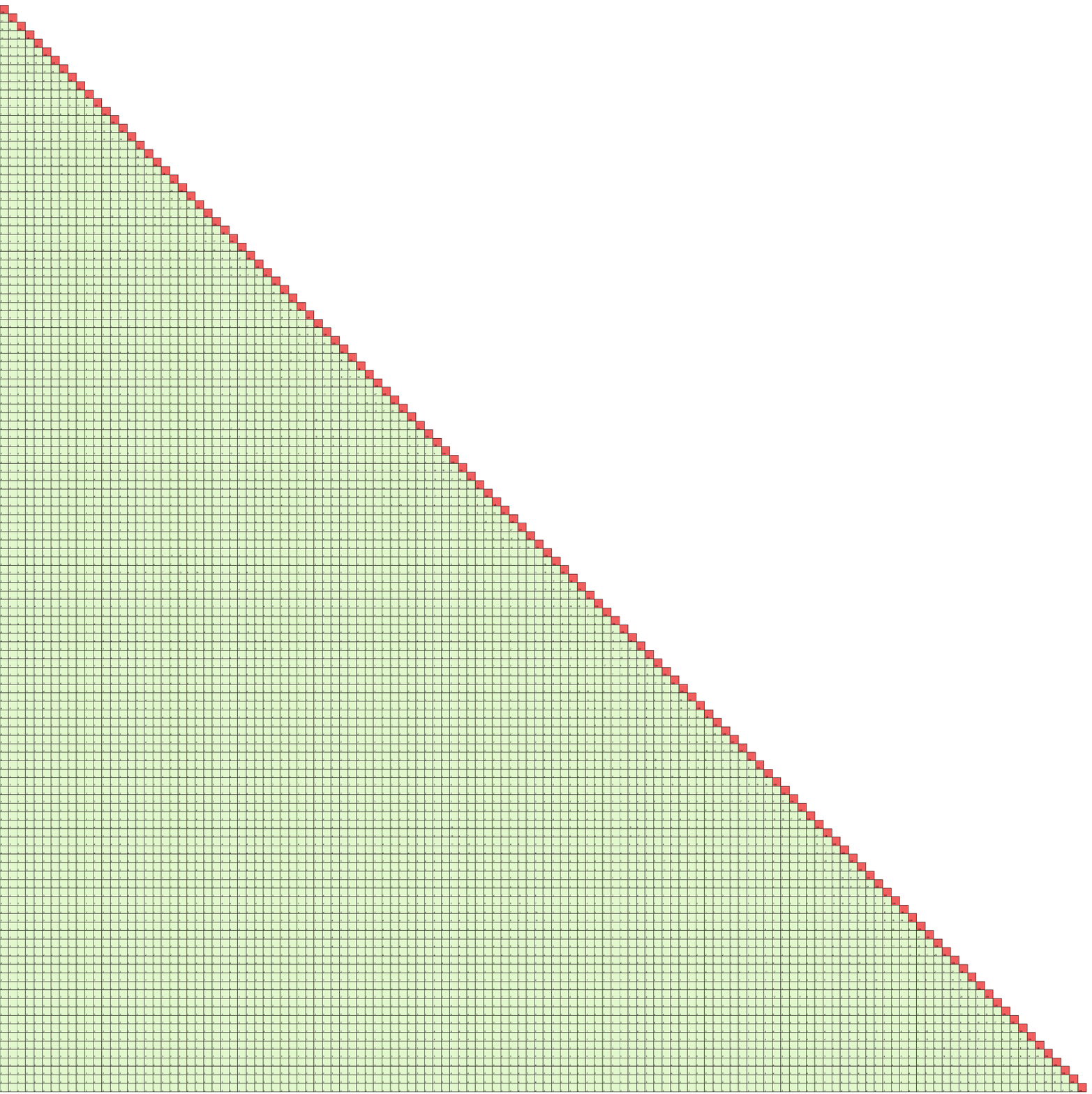}}} \\
  \subfloat[HODLR \& Reorder]{\label{fig:n1}{\includegraphics[width=0.29\textwidth]{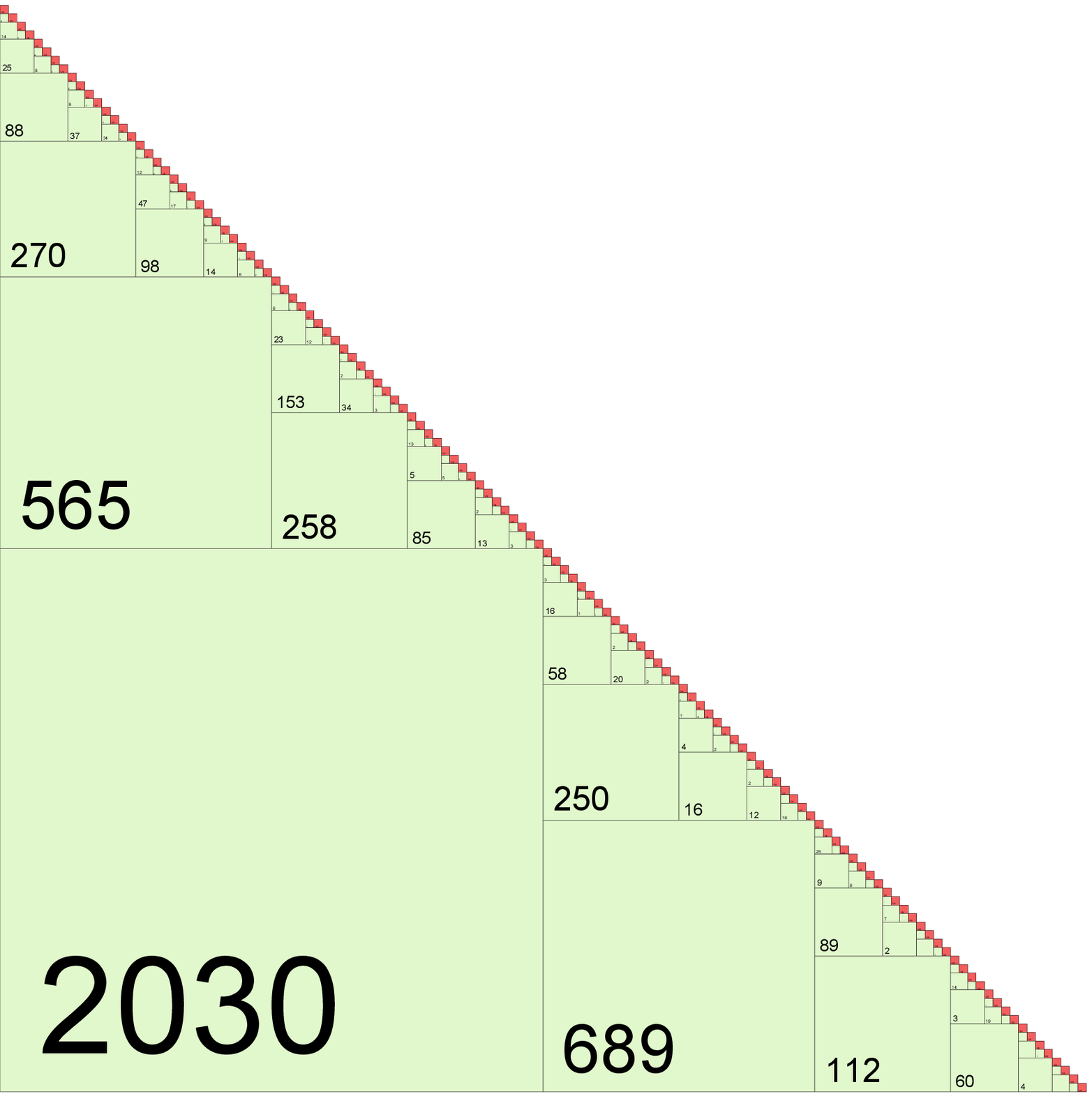}}}
  \hspace*{\fill}
  \subfloat[Standard \& Reorder]{\label{fig:n2}{\includegraphics[width=0.29\textwidth]{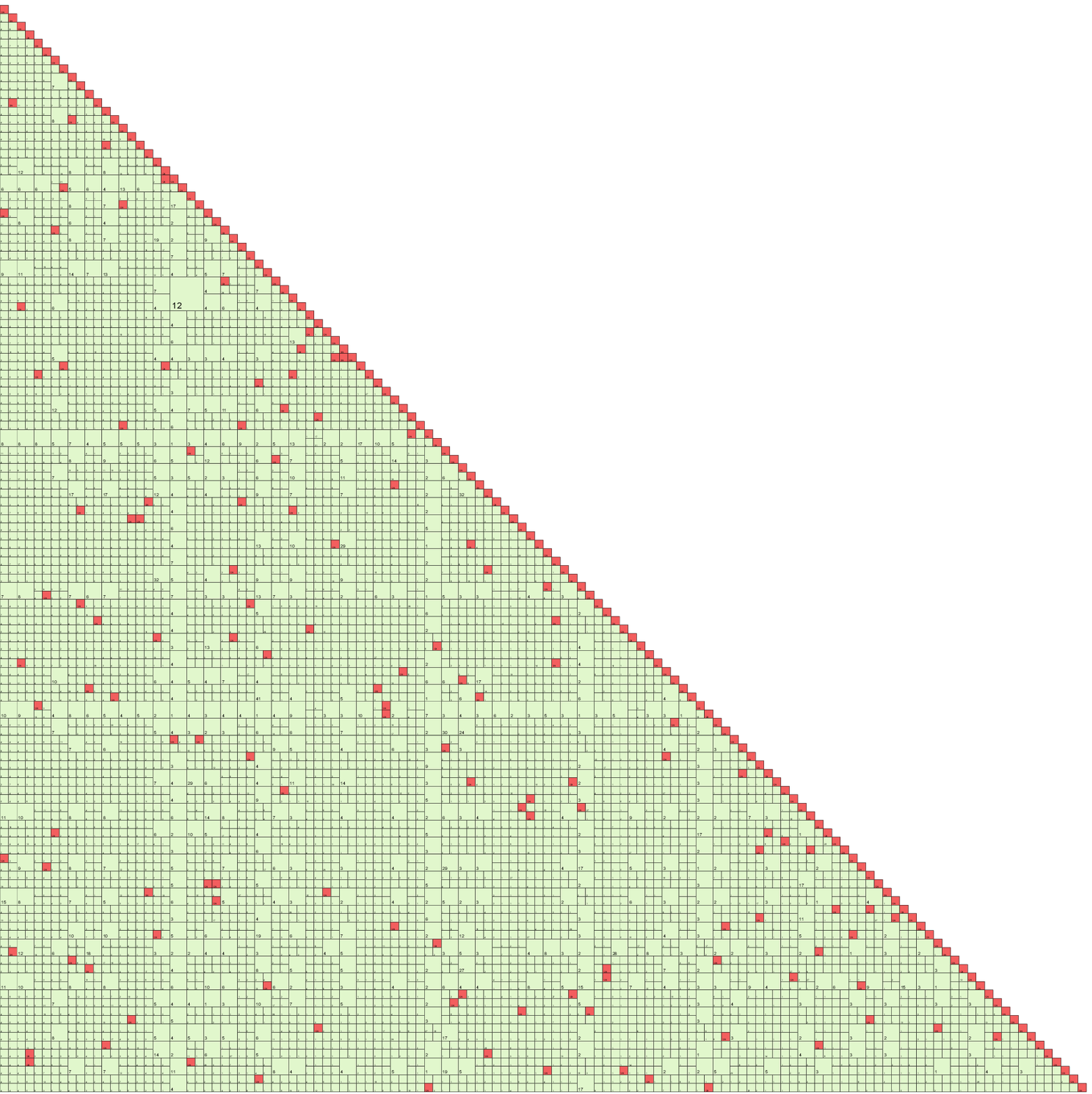}}}
  \hspace*{\fill}
  \subfloat[TLR \& Reorder]{\label{fig:n2}{\includegraphics[width=0.29\textwidth]{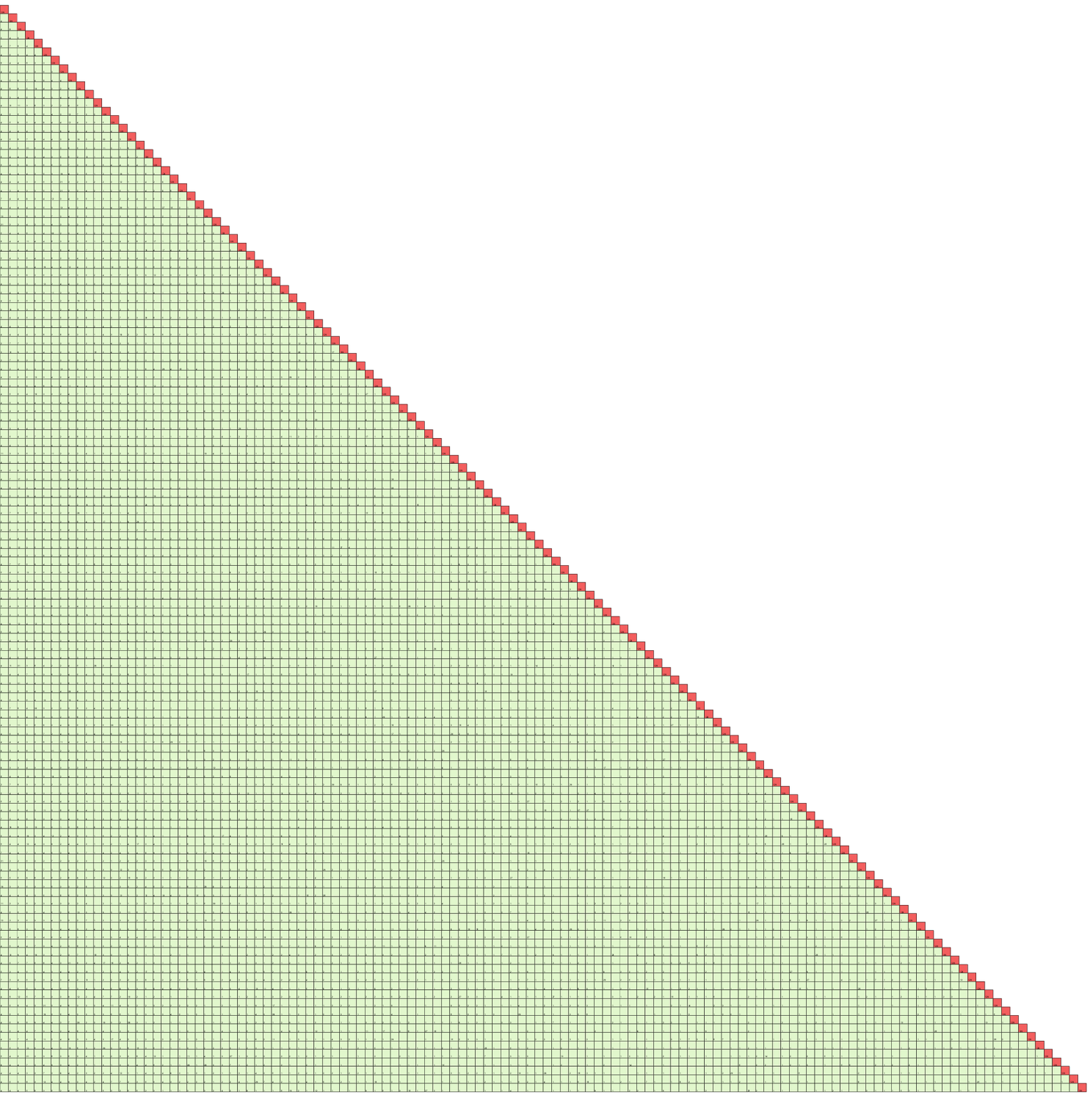}}} \\
  \caption{Partition and local ranks of a 16{,}384-dimensional covariance matrix represented with three low-rank structures. Red denotes dense blocks and green is used for low-rank blocks whose ranks are the numbers inside. For each structure, two indexing methods are compared, namely a geometrical indexing (Geom) and the block reordering (Reorder). HODLR and STD are the hierarchical structures under the weak and the standard admissibility conditions, respectively. TLR is the tile-low-rank structure. The covariance matrix is built with a grid in the unit square and the exponential covariance kernel, $\exp(- {\|\bh\|} / 0.3)$, where $\bh$ is the vector connecting two locations.}
  \label{fig:lc_rk_part}
\end{figure}

To show the TLR structure has good compatibility with the block-reordering scheme introduced in \cite{CGKT2018hcmvn}, we consider an MVN problem whose integration limits are independent from the geometry. The geometry is a $128 \times 128$ grid in the unit square whose locations are initially indexed with the geometrical clustering method provided in HLIBpro v2.8 \citep{borm2003introduction, kriemann2005parallel, grasedyck2008parallel}. The partitioning of each cluster is cardinality balanced, i.e., the two child clusters have equal number of indices and the minimum cluster size is set to 128. This geometrical indexing is used as the benchmark for measuring the efficiency of three low-rank structures discussed in \Cref{fig:lc_rk_part}. Assuming that the integration limits are independent and identically distributed, the block reordering is equivalent to shuffling the 128 clusters previously computed. A cluster-wise shuffle of the first indexing is used for measuring the compatibility of the low-rank representations with the integration-oriented ordering. \Cref{fig:lc_rk_part} describes the approximation of the Cholesky factors of the two covariance matrices using the hierarchical structures under the weak and the standard admissibility conditions as well as the TLR structure while \Cref{tbl:chol_time_mem} lists the corresponding time costs and memory footprints of the factorization.

\begin{table}
  \centering
  \caption{Cholesky factorization times and memory footprints for a 16{,}384-dimensional covariance matrix under a geometrical indexing (Geom) and block reordering (Reorder). HODLR and STD are the hierarchical structures under the weak and the standard admissibility conditions, respectively. TLR is the tile-low-rank structure. The covariance matrix is built with a grid in the unit square and the exponential covariance kernel, $\exp(- {\|\bh\|} / 0.3)$, where $\bh$ is the vector connecting two locations.}
  \label{tbl:chol_time_mem}
  \begin{tabular}{l  r  r  r  r}
    \hline \smallskip
    & factorize (Geom) & factorize (Reorder) & memory (Geom)  & memory (Reorder)\\
    \hline \smallskip
    HODLR & $5.7s$ & $46.0s$ & 102MB & 409MB\\
    STD & $0.6s$ & $2.8s$ & 60MB & 82MB\\
    TLR & $4.5s$ & $2.9s$ & 88MB & 67MB\\
  \end{tabular}
\end{table}

We focus on the Cholesky factor instead of the covariance matrix because the former's memory footprint has a linear relationship with the cost per sample in the computation of MVN probabilities. For each low-rank structure, the covariance matrix is constructed and factorized using the fixed-precision truncation with an absolute error of $10^{-4}$. It is worth mentioning that this truncation accuracy is unnecessarily high for the TLR structure and the hierarchical structure under the standard admissibility condition while necessary for the Cholesky factorization of the hierarchical structure under the weak admissibility condition (HODLR). Therefore in practice, we expect higher efficiency than what \Cref{fig:lc_rk_part} indicates when using the former two low-rank structures. In contrast, the fixed-rank truncation is less suitable for our purpose because the local ranks have large variability. Choosing a high fixed rank would waste memory while using a low fixed rank could cause factorization failure. Furthermore, the fixed-precision truncation with a relative error is empirically found to be less efficient than that with an absolute error, generating higher memory footprints at the accuracy threshold for a successful factorization.

The selected covariance kernel, $\exp({- \|\bh\|} / 0.3)$, has a range parameter $\beta = 0.3$ and an effective range of $0.9$, indicating a strong correlation in the unit square. Overall, stronger correlation presents more challenge in terms of Cholesky factorization because the covariance matrix becomes more numerically singular. The exponential kernel corresponds to the Mat\'ern kernel with a smoothness parameter $\nu = 0.5$ and the correlation for small $\|\bh\|$ (relative to $\beta$) increases asymptotically to one when $\nu$ increases, for which the singularity of the Mat\'ern kernel generally grows with $\nu$ if the given locations are in a compact domain. Therefore, smooth kernels are prone to having singularity issues, in which case either increasing the truncation accuracy or adding a nugget effect can enhance the factorization.

The admissibility condition measures the ratio of the distance between two clusters to a weighted average of their diameters. Hence, a big ratio indicates good separability and a small local rank, whereas the HODLR structure has large blocks in the low-rank format whose ratios are close or even equal to zero, leading to high memory footprint. Additionally, its increase is also the largest when switching from the geometrical indexing to the block reordering because a shuffle to the leaf clusters has stronger impact on the blocks whose row and column clusters consist of a larger number of leaf clusters. Contrarily, the memory footprint of using the TLR structure even decreases when switching to the block reordering. This may appear surprising but an explanation is given from two perspectives:
\begin{enumerate}
  \item The overall local ranks in the TLR covariance matrix do not change. The block reordering shuffles the leaf clusters, which only rearranges the off-diagonal blocks since the block size in the TLR representation is equal to the size of leaf clusters. 
  \item Secondly, the magnitude of the Schur complement decreases faster with the block column index under the block reordering. The Schur complement at block column $i$, $\bSigma_{i:r, i:r} - \bSigma_{i:r, 1:i-1} \bSigma_{1:i-1, 1:i-1}^{-1} \bSigma_{i:r, 1:i-1}^{\top}$, is the conditional covariance matrix for variables in clusters $i$ to $r$ given variables in clusters $1$ to $i-1$. We argue that for general spatial kernels, the conditional covariance matrix tends to have smaller magnitude, e.g., the Frobenius norm, when the locations of the conditioning variables are more scattered. This is treated as a heuristic without proof since it is not the focus of the paper. 
\end{enumerate}
Therefore, the block reordering maintains the local ranks of the TLR covariance matrix while increasing the magnitude decay of its Cholesky factor, which collectively reduce the memory footprint of the TLR Cholesky factor under the fixed-precision truncation with an absolute error. The combination of geometrical indexing and the standard admissibility condition produces the smallest memory footprint and factorization time. However, it is unable to benefit from an improved convergence rate. Under the block reordering, the standard admissibility condition produces a fine block partition that approaches the TLR structure and has a higher memory footprint than the TLR structure. The reason is that not satisfying the standard admissibility condition does not guarantee that the block's dense representation is more economical than its low-rank representation and in fact, the opposite is true for many dense blocks.

We conclude that under the block reordering, the TLR structure has the highest efficiency among the three low-rank structures discussed above and is only slightly less efficient than the combination of geometrical indexing and the standard admissibility condition. Additionally, the TLR structure is conceptually simpler than the other two and hence, more likely to benefit from parallel hardware architectures.

\subsection{Reordering schemes and TLR factorizations}
\label{subsec:impact_ranks}

The block-reordering scheme was proposed in \cite{CGKT2018hcmvn} and shown to improve the estimation accuracy of the conditioning method at a lower cost than the univariate or bivariate reordering scheme introduced in \cite{trinh2015bivariate}. In this paper, we improve the original block-reordering scheme by ordering the clusters of variables iteratively. The new iterative block reordering, similar to the block version of the univariate reordering scheme in \cite{trinh2015bivariate}, enjoys a higher convergence rate and produces the Cholesky factor simultaneously.

\Cref{alg:border} describes the original block-reordering scheme proposed in \cite{CGKT2018hcmvn} while \Cref{alg:rborder} is the iterative version that produces the Cholesky factor. We use $\bSigma_{i,j}$ to represent the $(i,j)$-th size-$m$ block of $\bSigma$. Similar notations are also used for $\ba$ and $\bb$. The symbol $\rightleftharpoons$ indicates the switching of coefficients, rows, or columns. Variables can be overwritten by themselves after computations for performance benefits. When $i \neq j$, $\bSigma_{i,j}$ is stored in the low-rank format. The blue lines in \Cref{alg:rborder} mark the matrix operations that are also in the TLR Cholesky factorization \citep{akbudak2017tile}. If we ignore the cost for steps~5 and 9, the complexity of \Cref{alg:rborder} is the same as the TLR Cholesky factorization. Although the complexity for accurately computing $\Phi_m$ and the truncated expectations is high, the univariate conditioning method \citep{trinh2015bivariate}, with a complexity of $O(m^3)$, can provide an estimate for both that is indicative enough. \Cref{alg:border} ignores the correlation between the size-$m$ clusters and also uses the univariate conditioning method for approximating $\Phi_m$. Therefore, the block-reordering scheme has a total complexity of $O(nm^2)$ but requires a succeeding Cholesky factorization while the iterative block reordering has additional complexity of $O(n^2m)$ over the TLR Cholesky factorization but produces the Cholesky factor simultaneously.
\begin{algorithm}
\caption{Block reordering}
\label{alg:border}
\begin{algorithmic}[1]
\State{bodr($\bSigma, \ba, \bb, m$)}
\State $r = n/m$
\For{$j = 1:r$} 
\State $\mathbf{p}[l] \approx \Phi_m(\ba_l,\bb_l;\bSigma_{l,l})$
\EndFor
\For{$j = 1:r$} 
\State $\tilde{j} = \mbox{argmin}_l(\mathbf{p}[l]), l = j,\ldots,r$
\State $\mathbf{p}[j \rightleftharpoons \tilde{j}]$ and block-wise $\bSigma[j \rightleftharpoons \tilde{j}, j \rightleftharpoons \tilde{j}]$, $\ba[j \rightleftharpoons \tilde{j}]$, $\bb[j \rightleftharpoons \tilde{j}]$
\EndFor
\end{algorithmic}
\end{algorithm}

\begin{algorithm}[!ht]
\caption{Block reordering during Cholesky factorization}
\label{alg:rborder}
\begin{algorithmic}[1]
\State{rbodr($\bSigma, \ba , \bb , m$)}
\State $r = n/m$
\For{$j = 1:r$} 
\For{$l = j:r$}
\State $\mathbf{p}[l] \approx \Phi_m(\ba_l,\bb_l;\bSigma_{l,l})$
\EndFor
\State $\tilde{j} = \mbox{argmin}_l(\mathbf{p}[l]), l = j,\ldots,r$
\State Block-wise $\bSigma[j \rightleftharpoons \tilde{j}, j \rightleftharpoons \tilde{j}]$, $\ba[j \rightleftharpoons \tilde{j}]$, $\bb[j \rightleftharpoons \tilde{j}]$
\State $\by_j \approx E_m[\bY|\bY \sim N_m(\0,\bSigma_{j,j}), \bY \in (\ba_j , \bb_j)]$
\State {\color{blue} $\bSigma_{j,j} = \mbox{Cholesky}(\bSigma_{j,j})$}
\For{$i = j+1:r$}
\State {\color{blue}$\bSigma_{i,j} = \bSigma_{i,j} \odot \bSigma_{j,j}^{-\top}$}
\State $\ba_i = \ba_i - \bSigma_{i,j} \odot \by_j$, $\bb_i = \bb_i - \bSigma_{i,j} \odot \by_j$
\EndFor
\For{$j_1 = j+1:r$}
\For{$i_1 = j+1:r$}
\State {\color{blue}$\bSigma_{i_1,j_1} = \bSigma_{i_1,j_1} \ominus \bSigma_{i_1,j} \odot \bSigma_{j_1,j}^{\top}$}
\EndFor
\EndFor
\EndFor
\end{algorithmic}
\end{algorithm}

The truncated operations, $\odot$ and $\ominus$, indicate that matrix product and matrix subtraction are followed by truncation to smaller ranks while maintaining the required accuracy, given that the block-wise ranks may generally expand as a result of the operations. Here, $\bSigma_{i_1,j} \odot \bSigma_{j_1,j}^{\top}$ and $\bSigma_{i,j} \odot \bSigma_{j,j}^{-\top}$ have complexities of $O(mk^2)$ and $O(m^2k)$ respectively, where $m$ is the tile size and $k$ is the local rank. The $\ominus$ operation uses ACA truncated at an absolute tolerance to keep the result low-rank. For the studies in \Cref{sec:simulation} and \Cref{sec:application}, we set the tolerance to $10^{-5}$. Prior to the TLR Cholesky factorization, we construct the TLR covariance matrix with ACA given the covariance kernel, the underlying geometry and the indices of variables. Therefore, the total memory needed for computing MVN and MVT probabilities is $O(kn^2/m)$.


\subsection{Preconditioned TLR QMC algorithms}
\label{subsec:tlr_struct}

\Cref{alg:tlrmvn} and \Cref{alg:tlrmvt} describe the TLR versions of the `mvn' and `mvt' algorithms. To distinguish them from the dense `mvn' and `mvt' algorithms, we expand the storage structure of $\bL$, the TLR Cholesky factor, as the interface of the TLR algorithms. The definitions of $\bB_i$, $\bU_{i, j}$, and $\bV_{i, j}$ are shown in \Cref{fig:structure}.

Similar to \Cref{alg:rborder},
\begin{algorithm}
\caption{TLR QMC for MVN probabilities}
\label{alg:tlrmvn}
\begin{algorithmic}[1]
\State{tlrmvn($\bB, \bU, \bV, \ba, \bb, \bw$)}
\State $\by \gets \mathbf{0}$, and $P \gets 1$
\For{$i = 1:r$} 
\If{$i > 1$} 
\For{$j = i:r$}
\State $\bm{\Delta} = \bU_{j,i-1}(\bV_{j,i-1}^T\by_{i-1})$
\State $\ba_j = \ba_j - \bm{\Delta}$, $\bb_j = \bb_j - \bm{\Delta}$
\EndFor
\EndIf
\State $(P',\by_i) \gets \mbox{MVN}(\bB_i,\ba_i,\bb_i,\bw_i)$
\State $P \gets P \cdot  P'$
\EndFor
\State \Return $P$
\end{algorithmic}
\end{algorithm}
\begin{algorithm}
\caption{TLR QMC for MVT probabilities}
\label{alg:tlrmvt}
\begin{algorithmic}[1]
\State{tlrmvt($\bB, \bU, \bV, \ba, \bb, \nu, w_0 , \bw$)}
\State $\ba' \gets \frac{\chi^{-1}_\nu(w_0)}{\sqrt{\nu}} \ba$, $\bb' \gets \frac{\chi^{-1}_\nu(w_0)}{\sqrt{\nu}} \bb$
\State \Return $\mbox{TLRMVN}(\bB, \bU, \bV, \ba',\bb',\bw)$
\end{algorithmic}
\end{algorithm}
we use subscripts to represent the size-$m$ segment of $\ba$, $\bb$, $\by$, and $\bw$. The two algorithms compute the integrand given one sample $\bw$ in the $n$-dimensional unit hypercube. In our implementation, the Richtmyer rule \citep{richtmyer1951evaluation}, recommended by \cite{genz2009computation}, is employed for choosing $\bw$. Here, `tlrmvn' is called by `tlrmvt', where the additional inputs, $\nu$ and $w_0$, bear the same meaning as those in \Cref{alg:mvt2}. The TLR structure reduces dense matrix-vector multiplication to low rank matrix-vector multiplication when factoring the correlation between size-$m$ clusters into the integration limits. The TLR structure reduces the complexity of matrix-vector multiplication, hence the cost per MC sample, at the step of block updating the integration limits (Lines~6 and 7 in \Cref{alg:tlrmvn}). The TLR QMC is a variant of the SOV algorithm from \cite{genz1992numerical} that belongs to the same category as the hierarchical QMC \citep{GKT2016hmvn}. \Cref{alg:tlrmvn} and \Cref{alg:tlrmvt} can be either preconditioned by the block reordering or the iterative block reordering. We examine the performance of the TLR QMC algorithms in \Cref{sec:simulation}. 


\section{Numerical Simulations}
\label{sec:simulation}
\begin{table}[b!]
{\footnotesize
\caption{Performance of the dense, hierarchical, and TLR methods for computing MVN/MVT probabilities. `mvn' and `mvt' are the dense QMC methods, `hmvn' and `hmvt' are the hierarchical QMC methods, `tlrmvn' and `tlrmvt' are the TLR QMC methods, `r' indicates the block reordering preconditioner, and `rr' indicates the iterative block reordering preconditioner. The upper row is the average relative estimation error and the lower row is the average computation time over $20$ replicates. The covariance matrix is generated from a 2D exponential kernel, $\exp(- \|\bh\| / \beta)$, where $\bh$ is the distance vector, based on $n$ locations on a perturbed grid in the unit square. The lower integration limits are $-\bm{\infty}$ and the upper limits are independently generated from $N(5.5, 1.25^2)$. The degrees of freedom $\nu$ for MVT probabilities are $10$. The QMC sample size for preconditioned methods is $10^{4}$ and $10^{3}$ for others.}
\label{tbl:simulations}
\begin{center}
\begin{tabular}{crrrrrrrrrr}
\multicolumn{11}{c}{$\beta = 0.3$ (strong correlation)} \\
\hline \noalign{\smallskip}
$n$ & mvn & hmvn & tlrmvn & rtlrmvn & rrtlrmvn & mvt & hmvt & tlrmvt & rtlrmvt & rrtlrmvt\\
\noalign{\smallskip}\hline\noalign{\smallskip}
$1024$ & 
\begin{tabular}{@{}r@{}} $0.5\%$  \\ \textit{2.5s}\end{tabular}& 
\begin{tabular}{@{}r@{}} $0.5\%$  \\ \textit{1.1s}\end{tabular}& 
\begin{tabular}{@{}r@{}} $0.5\%$  \\ \textit{1.7s}\end{tabular}& 
\begin{tabular}{@{}r@{}} $0.4\%$  \\ \textit{0.1s}\end{tabular}& 
\begin{tabular}{@{}r@{}} $0.4\%$  \\ \textit{0.1s}\end{tabular}& 
\begin{tabular}{@{}r@{}} $0.7\%$  \\ \textit{2.6s}\end{tabular}& 
\begin{tabular}{@{}r@{}} $1.5\%$  \\ \textit{1.1s}\end{tabular}& 
\begin{tabular}{@{}r@{}} $1.6\%$  \\ \textit{2.0s}\end{tabular}& 
\begin{tabular}{@{}r@{}} $1.7\%$  \\ \textit{0.2s}\end{tabular}& 
\begin{tabular}{@{}r@{}} $1.6\%$  \\ \textit{0.2s}\end{tabular}\\ 
$4096$ & 
\begin{tabular}{@{}r@{}} $1.1\%$  \\ \textit{42.4s}\end{tabular}& 
\begin{tabular}{@{}r@{}} $1.1\%$  \\ \textit{9.7s}\end{tabular}& 
\begin{tabular}{@{}r@{}} $1.0\%$  \\ \textit{14.1s}\end{tabular}& 
\begin{tabular}{@{}r@{}} $0.9\%$  \\ \textit{1.4s}\end{tabular}& 
\begin{tabular}{@{}r@{}} $1.0\%$  \\ \textit{1.3s}\end{tabular}& 
\begin{tabular}{@{}r@{}} $1.0\%$  \\ \textit{41.1s}\end{tabular}& 
\begin{tabular}{@{}r@{}} $1.6\%$  \\ \textit{7.9s}\end{tabular}& 
\begin{tabular}{@{}r@{}} $1.4\%$  \\ \textit{12.7s}\end{tabular}& 
\begin{tabular}{@{}r@{}} $1.1\%$  \\ \textit{1.2s}\end{tabular}& 
\begin{tabular}{@{}r@{}} $1.2\%$  \\ \textit{1.2s}\end{tabular}\\ 
$16384$ & 
\begin{tabular}{@{}r@{}} $2.2\%$  \\ \textit{1233.5s}\end{tabular}& 
\begin{tabular}{@{}r@{}} $2.3\%$  \\ \textit{56.2s}\end{tabular}& 
\begin{tabular}{@{}r@{}} $2.2\%$  \\ \textit{93.9s}\end{tabular}& 
\begin{tabular}{@{}r@{}} $2.0\%$  \\ \textit{8.2s}\end{tabular}& 
\begin{tabular}{@{}r@{}} $1.8\%$  \\ \textit{8.2s}\end{tabular}& 
\begin{tabular}{@{}r@{}} $4.5\%$  \\ \textit{1198.3s}\end{tabular}& 
\begin{tabular}{@{}r@{}} $4.3\%$  \\ \textit{49.9s}\end{tabular}& 
\begin{tabular}{@{}r@{}} $4.2\%$  \\ \textit{90.0s}\end{tabular}& 
\begin{tabular}{@{}r@{}} $2.8\%$  \\ \textit{7.8s}\end{tabular}& 
\begin{tabular}{@{}r@{}} $2.4\%$  \\ \textit{7.8s}\end{tabular}\\ 
$65536$ & 
\begin{tabular}{@{}r@{}} N.A.  \\ \textit{N.A.}\end{tabular}& 
\begin{tabular}{@{}r@{}} $5.8\%$  \\ \textit{309.5s}\end{tabular}& 
\begin{tabular}{@{}r@{}} $4.7\%$  \\ \textit{596.2s}\end{tabular}& 
\begin{tabular}{@{}r@{}} $3.5\%$  \\ \textit{50.8s}\end{tabular}& 
\begin{tabular}{@{}r@{}} $2.6\%$  \\ \textit{51.2s}\end{tabular}& 
\begin{tabular}{@{}r@{}} N.A.  \\ \textit{N.A.}\end{tabular}& 
\begin{tabular}{@{}r@{}} $13.5\%$  \\ \textit{294.9s}\end{tabular}& 
\begin{tabular}{@{}r@{}} $13.7\%$  \\ \textit{594.7s}\end{tabular}& 
\begin{tabular}{@{}r@{}} $6.3\%$  \\ \textit{49.9s}\end{tabular}& 
\begin{tabular}{@{}r@{}} $5.9\%$  \\ \textit{50.4s}\end{tabular}\\
\noalign{\smallskip}
\multicolumn{11}{c}{$\beta = 0.1$ (medium correlation)} \\
\hline \noalign{\smallskip}
$n$ & mvn & hmvn & tlrmvn & rtlrmvn & rrtlrmvn & mvt & hmvt & tlrmvt & rtlrmvt & rrtlrmvt\\
\noalign{\smallskip}\hline\noalign{\smallskip}
$1024$ & 
\begin{tabular}{@{}r@{}} $0.4\%$  \\ \textit{2.4s}\end{tabular}& 
\begin{tabular}{@{}r@{}} $0.4\%$  \\ \textit{1.0s}\end{tabular}& 
\begin{tabular}{@{}r@{}} $0.4\%$  \\ \textit{1.6s}\end{tabular}& 
\begin{tabular}{@{}r@{}} $0.4\%$  \\ \textit{0.1s}\end{tabular}& 
\begin{tabular}{@{}r@{}} $0.3\%$  \\ \textit{0.1s}\end{tabular}& 
\begin{tabular}{@{}r@{}} $0.5\%$  \\ \textit{2.4s}\end{tabular}& 
\begin{tabular}{@{}r@{}} $0.6\%$  \\ \textit{0.8s}\end{tabular}& 
\begin{tabular}{@{}r@{}} $0.5\%$  \\ \textit{1.6s}\end{tabular}& 
\begin{tabular}{@{}r@{}} $0.6\%$  \\ \textit{0.1s}\end{tabular}& 
\begin{tabular}{@{}r@{}} $0.7\%$  \\ \textit{0.1s}\end{tabular}\\ 
$4096$ & 
\begin{tabular}{@{}r@{}} $1.3\%$  \\ \textit{38.5s}\end{tabular}& 
\begin{tabular}{@{}r@{}} $1.3\%$  \\ \textit{5.2s}\end{tabular}& 
\begin{tabular}{@{}r@{}} $1.4\%$  \\ \textit{9.3s}\end{tabular}& 
\begin{tabular}{@{}r@{}} $1.0\%$  \\ \textit{1.0s}\end{tabular}& 
\begin{tabular}{@{}r@{}} $1.2\%$  \\ \textit{1.0s}\end{tabular}& 
\begin{tabular}{@{}r@{}} $1.1\%$  \\ \textit{38.4s}\end{tabular}& 
\begin{tabular}{@{}r@{}} $1.1\%$  \\ \textit{5.0s}\end{tabular}& 
\begin{tabular}{@{}r@{}} $1.3\%$  \\ \textit{9.9s}\end{tabular}& 
\begin{tabular}{@{}r@{}} $1.2\%$  \\ \textit{1.1s}\end{tabular}& 
\begin{tabular}{@{}r@{}} $1.2\%$  \\ \textit{1.0s}\end{tabular}\\ 
$16384$ & 
\begin{tabular}{@{}r@{}} $4.4\%$  \\ \textit{1188.2s}\end{tabular}& 
\begin{tabular}{@{}r@{}} $4.3\%$  \\ \textit{37.0s}\end{tabular}& 
\begin{tabular}{@{}r@{}} $4.1\%$  \\ \textit{78.7s}\end{tabular}& 
\begin{tabular}{@{}r@{}} $4.1\%$  \\ \textit{6.6s}\end{tabular}& 
\begin{tabular}{@{}r@{}} $3.5\%$  \\ \textit{6.6s}\end{tabular}& 
\begin{tabular}{@{}r@{}} $3.1\%$  \\ \textit{1176.8s}\end{tabular}& 
\begin{tabular}{@{}r@{}} $3.6\%$  \\ \textit{36.3s}\end{tabular}& 
\begin{tabular}{@{}r@{}} $2.8\%$  \\ \textit{79.6s}\end{tabular}& 
\begin{tabular}{@{}r@{}} $3.4\%$  \\ \textit{6.6s}\end{tabular}& 
\begin{tabular}{@{}r@{}} $3.2\%$  \\ \textit{6.6s}\end{tabular}\\ 
$65536$ & 
\begin{tabular}{@{}r@{}} N.A.  \\ \textit{N.A.}\end{tabular}& 
\begin{tabular}{@{}r@{}} $34.4\%$  \\ \textit{286.8s}\end{tabular}& 
\begin{tabular}{@{}r@{}} $31.1\%$  \\ \textit{562.9s}\end{tabular}& 
\begin{tabular}{@{}r@{}} $12.0\%$  \\ \textit{47.8s}\end{tabular}& 
\begin{tabular}{@{}r@{}} $11.6\%$  \\ \textit{49.5s}\end{tabular}& 
\begin{tabular}{@{}r@{}} N.A.  \\ \textit{N.A.}\end{tabular}& 
\begin{tabular}{@{}r@{}} $17.2\%$  \\ \textit{254.7s}\end{tabular}& 
\begin{tabular}{@{}r@{}} $17.0\%$  \\ \textit{531.6s}\end{tabular}& 
\begin{tabular}{@{}r@{}} $7.7\%$  \\ \textit{44.9s}\end{tabular}& 
\begin{tabular}{@{}r@{}} $7.4\%$  \\ \textit{45.4s}\end{tabular}\\
\noalign{\smallskip}
\multicolumn{11}{c}{$\beta = 0.03$ (weak correlation)} \\
\hline \noalign{\smallskip}
$n$ & mvn & hmvn & tlrmvn & rtlrmvn & rrtlrmvn & mvt & hmvt & tlrmvt & rtlrmvt & rrtlrmvt\\
\noalign{\smallskip}\hline\noalign{\smallskip}
$1024$ & 
\begin{tabular}{@{}r@{}} $0.1\%$  \\ \textit{2.4s}\end{tabular}& 
\begin{tabular}{@{}r@{}} $0.2\%$  \\ \textit{0.9s}\end{tabular}& 
\begin{tabular}{@{}r@{}} $0.2\%$  \\ \textit{1.5s}\end{tabular}& 
\begin{tabular}{@{}r@{}} $0.1\%$  \\ \textit{0.1s}\end{tabular}& 
\begin{tabular}{@{}r@{}} $0.1\%$  \\ \textit{0.1s}\end{tabular}& 
\begin{tabular}{@{}r@{}} $0.2\%$  \\ \textit{2.3s}\end{tabular}& 
\begin{tabular}{@{}r@{}} $0.2\%$  \\ \textit{0.8s}\end{tabular}& 
\begin{tabular}{@{}r@{}} $0.2\%$  \\ \textit{1.6s}\end{tabular}& 
\begin{tabular}{@{}r@{}} $0.4\%$  \\ \textit{0.1s}\end{tabular}& 
\begin{tabular}{@{}r@{}} $0.4\%$  \\ \textit{0.1s}\end{tabular}\\ 
$4096$ & 
\begin{tabular}{@{}r@{}} $0.7\%$  \\ \textit{38.1s}\end{tabular}& 
\begin{tabular}{@{}r@{}} $0.7\%$  \\ \textit{4.9s}\end{tabular}& 
\begin{tabular}{@{}r@{}} $0.8\%$  \\ \textit{8.9s}\end{tabular}& 
\begin{tabular}{@{}r@{}} $0.5\%$  \\ \textit{0.9s}\end{tabular}& 
\begin{tabular}{@{}r@{}} $0.5\%$  \\ \textit{0.9s}\end{tabular}& 
\begin{tabular}{@{}r@{}} $0.7\%$  \\ \textit{37.7s}\end{tabular}& 
\begin{tabular}{@{}r@{}} $0.7\%$  \\ \textit{4.4s}\end{tabular}& 
\begin{tabular}{@{}r@{}} $0.8\%$  \\ \textit{9.1s}\end{tabular}& 
\begin{tabular}{@{}r@{}} $0.7\%$  \\ \textit{0.9s}\end{tabular}& 
\begin{tabular}{@{}r@{}} $0.8\%$  \\ \textit{0.9s}\end{tabular}\\ 
$16384$ & 
\begin{tabular}{@{}r@{}} $3.5\%$  \\ \textit{1118.5s}\end{tabular}& 
\begin{tabular}{@{}r@{}} $3.6\%$  \\ \textit{29.4s}\end{tabular}& 
\begin{tabular}{@{}r@{}} $4.0\%$  \\ \textit{64.5s}\end{tabular}& 
\begin{tabular}{@{}r@{}} $2.8\%$  \\ \textit{5.4s}\end{tabular}& 
\begin{tabular}{@{}r@{}} $2.4\%$  \\ \textit{5.4s}\end{tabular}& 
\begin{tabular}{@{}r@{}} $2.6\%$  \\ \textit{1097.4s}\end{tabular}& 
\begin{tabular}{@{}r@{}} $2.4\%$  \\ \textit{27.5s}\end{tabular}& 
\begin{tabular}{@{}r@{}} $2.3\%$  \\ \textit{65.0s}\end{tabular}& 
\begin{tabular}{@{}r@{}} $1.7\%$  \\ \textit{5.5s}\end{tabular}& 
\begin{tabular}{@{}r@{}} $1.6\%$  \\ \textit{5.4s}\end{tabular}\\ 
$65536$ & 
\begin{tabular}{@{}r@{}} N.A.  \\ \textit{N.A.}\end{tabular}& 
\begin{tabular}{@{}r@{}} $67.0\%$  \\ \textit{201.1s}\end{tabular}& 
\begin{tabular}{@{}r@{}} $75.3\%$  \\ \textit{450.2s}\end{tabular}& 
\begin{tabular}{@{}r@{}} $13.7\%$  \\ \textit{37.9s}\end{tabular}& 
\begin{tabular}{@{}r@{}} $14.2\%$  \\ \textit{37.9s}\end{tabular}& 
\begin{tabular}{@{}r@{}} N.A.  \\ \textit{N.A.}\end{tabular}& 
\begin{tabular}{@{}r@{}} $13.5\%$  \\ \textit{203.6s}\end{tabular}& 
\begin{tabular}{@{}r@{}} $13.5\%$  \\ \textit{459.8s}\end{tabular}& 
\begin{tabular}{@{}r@{}} $8.8\%$  \\ \textit{39.0s}\end{tabular}& 
\begin{tabular}{@{}r@{}} $8.1\%$  \\ \textit{38.6s}\end{tabular}\\
\end{tabular} 
\end{center}
}
\end{table}
\Cref{tbl:simulations} summarizes the performance of the dense \citep{genz1992numerical}, the hierarchical \citep{GKT2016hmvn}, and the TLR QMC methods for computing  MVN and MVT probabilities. Methods are assessed over $20$ simulated problems for each combination of problem dimension $n$ and correlation strength $\beta$. $\beta = 0.3$, $0.1$, and $0.03$ correspond to the effective ranges of $0.90$, $0.30$, and $0.09$, respectively, representing strong, medium, and weak correlation strengths in the unit square. The tile size $m$ for the TLR QMC methods is set as $\sqrt{n}$ for the optimal complexity per sample, $O(n^{3/2})$. Overall, smaller blocks are more easily represented in the low-rank format, i.e., having lower local ranks but a partition that is too fine may compromise the memory savings while bigger blocks can lead to higher local ranks and higher memory footprint for the diagonal blocks. For ease of comparison, the diagonal block size for the hierarchical QMC methods is also set as $\sqrt{n}$. We apply a fixed-precision truncation of $10^{-4}$ to $\beta = 0.3$ and of $10^{-3}$ to the other correlation strengths to guarantee the success of Cholesky factorization while enhancing computation efficiency. It is worth mentioning that the error caused by the truncation to a lower rank is typically invisible compared with that from the Monte Carlo integration since the difference between the estimates of the same MVN/MVT problem in \Cref{tbl:simulations} is well explained by their Monte Carlo standard errors. The sample size is $N = 10^4$ for the methods without any preconditioner and $N = 10^3$ for the four preconditioned methods to highlight their computation times for reaching the same accuracy. The listed time in \Cref{tbl:simulations} covers only the integration algorithm, not including the construction of the covariance matrix, the block reordering, and the Cholesky factorization. The computation time for Cholesky factorization is indicated in \Cref{tbl:chol_time_mem}, generally smaller than that of the Monte Carlo integration with $10^{4}$ samples by more than one order of magnitude. The iterative block reordering has the same order of complexity, $O(n^{5/2})$, as the (TLR) Cholesky factorization when $m = \sqrt{n}$ while the block reordering has the same complexity as the first iteration of the iterative block reordering. The complexity for constructing the covariance matrix is negligible, one order smaller than the Cholesky factorization. The highest dimension in our experiment is $2^{16}$. Considerations for higher dimensions include the truncation precision required for Cholesky factorization and the sample size needed to reach the desired accuracy. 

\begin{figure}[t!]
\centering
\subfloat[MVN]{\label{fig:rel_err_mvn}
\includegraphics[width=0.43\textwidth]{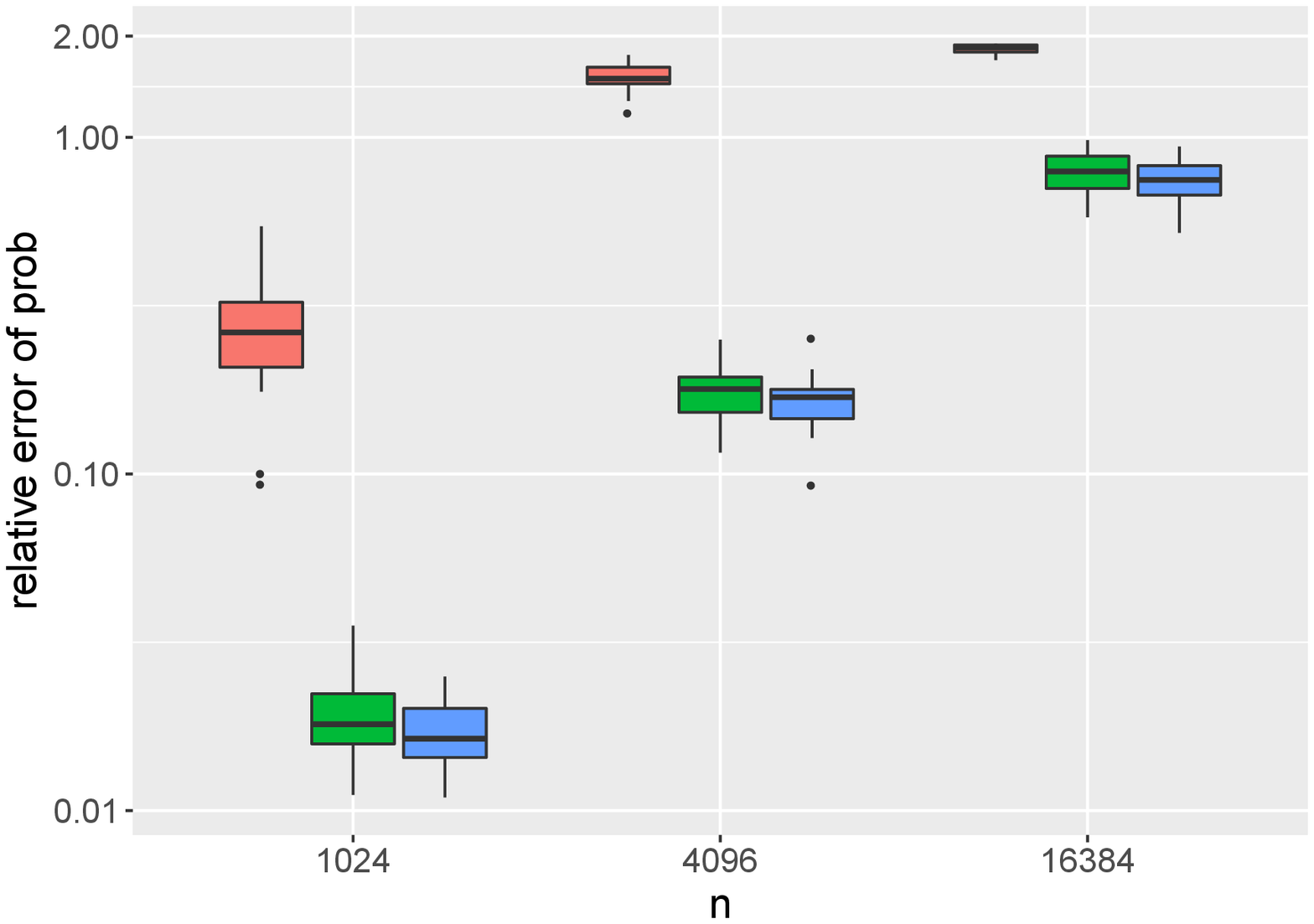}} 
\subfloat[MVT]{\label{fig:rel_err_mvt}
\includegraphics[width=0.43\textwidth]{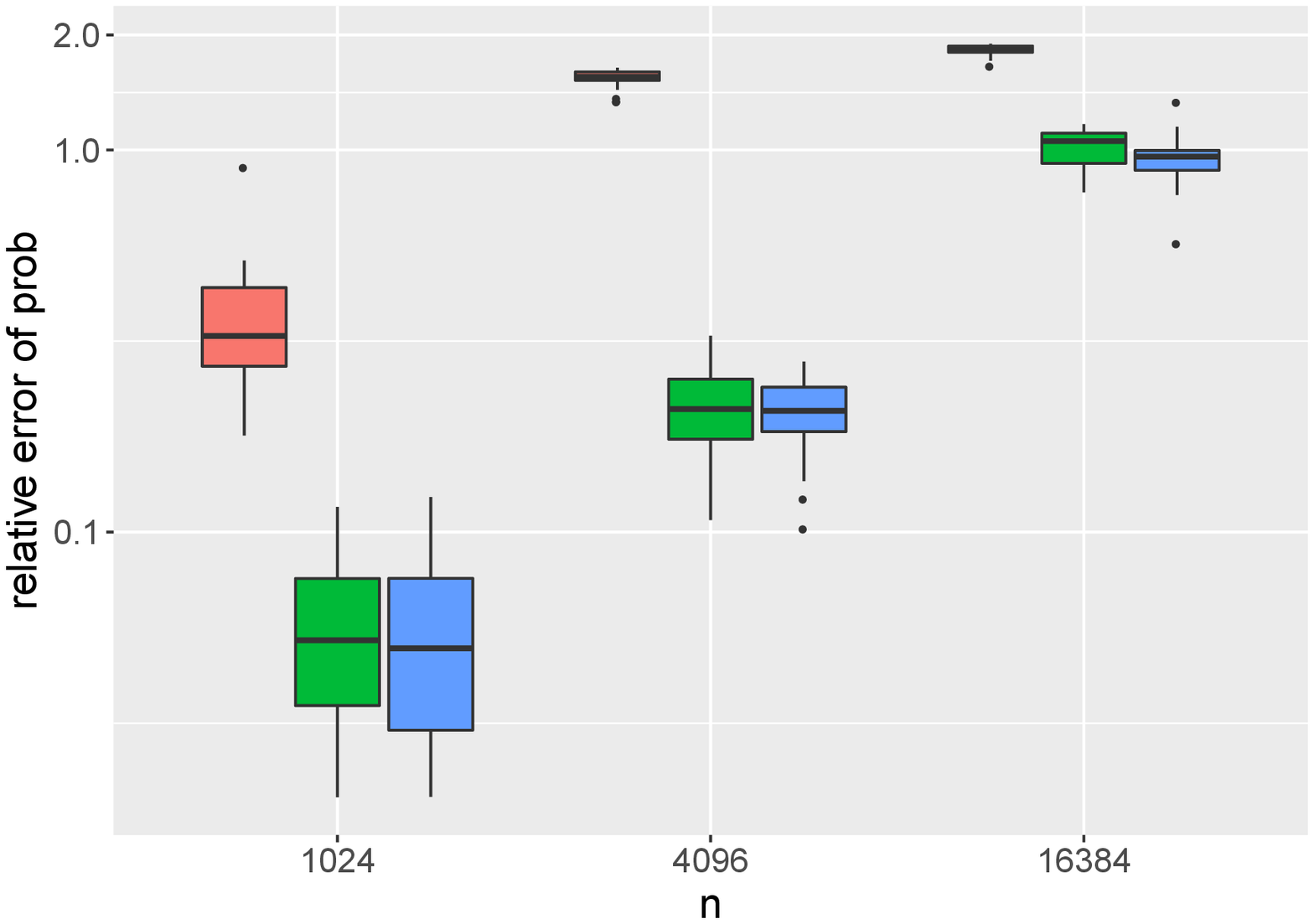}} \\
\subfloat[log-MVN]{\label{fig:log_err_mvn}
\includegraphics[width=0.43\textwidth]{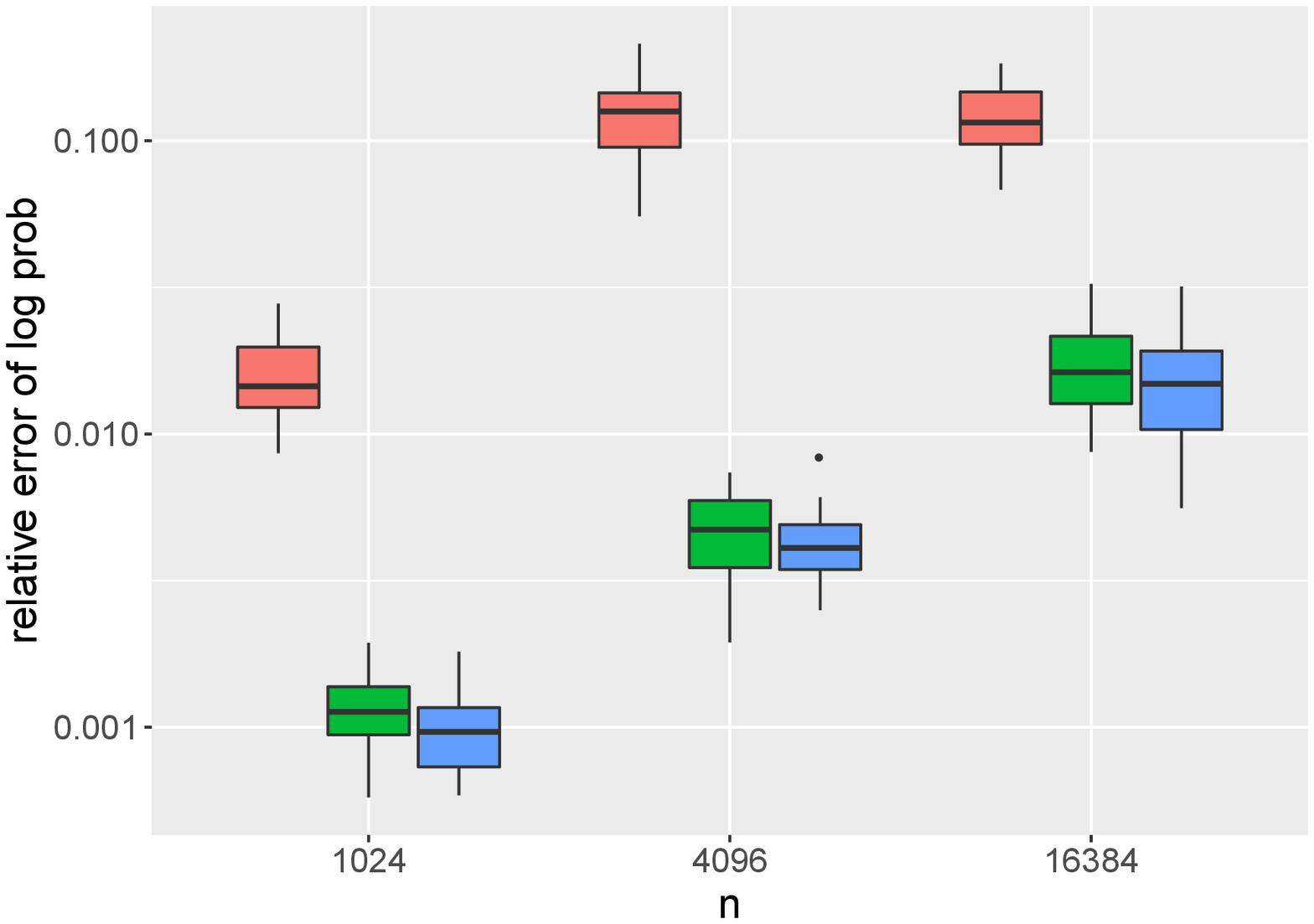}}
\subfloat[log-MVT]{\label{fig:log_err_mvt}
\includegraphics[width=0.43\textwidth]{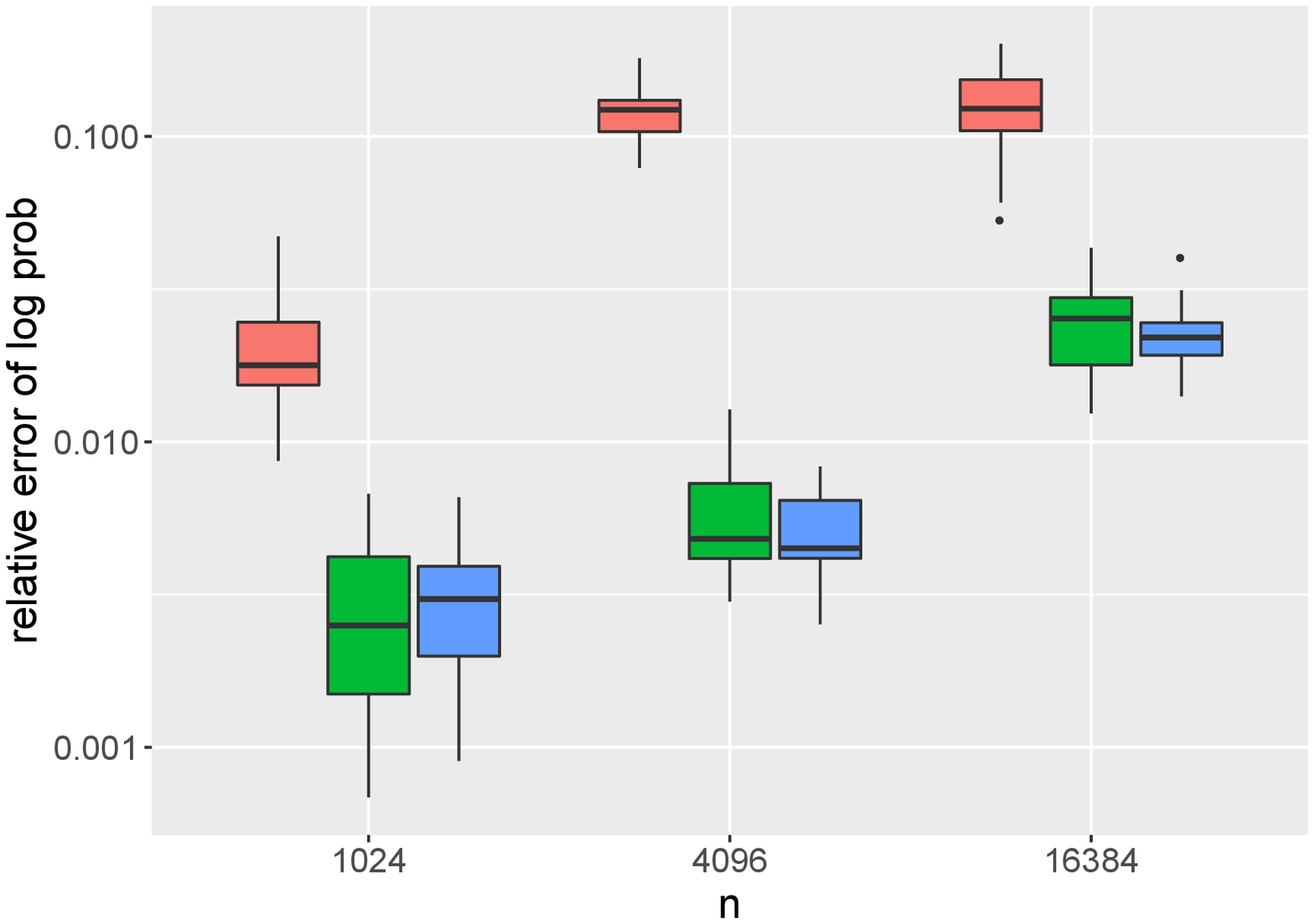}}
\caption{The relative error for probabilities and log-probabilities. For each $n$, the three boxplots, from left to right, correspond to the TLR method, the TLR with the block reordering method, and the TLR with the iterative block reordering method. The relative error for log-probabilities is based on $10$ estimations of the same replicate. Each boxplot consists of $20$ replicates. The covariance matrix is generated from a 2D exponential kernel, $\exp(- \|\bh\| / \beta)$, where $\bh$ is the distance vector, based on $n$ locations on a perturbed grid in the unit square. The lower integration limits are $-\bm{\infty}$ and the upper limits are independently generated from $N(4.0, 1.5^2)$. The degrees of freedom $\nu$ for MVT probabilities are $10$. The QMC sample size is $10^{4}$.}
\label{fig:rel_err}
\end{figure}

Considering only the methods without any preconditioner, the low-rank methods are more scalable than the dense methods and the time difference already reaches two orders of magnitude when $n = 16{,}384$. The hierarchical methods are more efficient than the TLR methods although \Cref{tbl:chol_time_mem} indicates slightly higher memory footprint for the hierarchical Cholesky factor under the weak admissibility condition. This is because the hierarchical methods involve fewer but larger matrix-matrix multiplications compared with the TLR methods, beneficial from modern optimized Level 3 BLAS functions. However, this weakness of the TLR methods is amenable to the future optimized multiplication routine between TLR and dense matrices. Furthermore, hierarchical methods are more sensitive to strong correlation, demanding higher truncation precision and hence, higher local ranks when the covariance matrix approaches singularity. After using the (iterative) block reordering preconditioner, the TLR methods reach even lower estimation error with one-tenth of the previous sample size, shortening the computation time by up to one order of magnitude compared with the hierarchical methods. Additionally, the iterative block reordering has slightly bigger improvement on the Monte Carlo convergence rate than the non-iterative one, which is seen more clearly in \Cref{fig:rel_err}. It is worth mentioning that the block reorderings are more effective when the MVN/MVT problem is more asymmetric. An extreme scenario where reorderings become ineffective is that the correlation is constant and the integration limits are the same across the variables. 

MVN and MVT probabilities much smaller than those in \Cref{tbl:simulations} may appear in high-dimensional applications \citep{botev2017normal}. For example, the model and data used in \Cref{subsec:est_real} produce a likelihood smaller than $10^{-40}$. Overall, the convergence rate decreases if the integration region is pushed towards the tail while keeping the covariance structure unchanged, for which standard scientific workstations may ultimately fail to reach the desired accuracy using a reasonably large sample size $N$. For example, when $n = 65{,}536$ and $\beta = 0.03$, the MVN methods without block reordering are unable to control the relative error with $10^4$ QMC samples, rendering even the most significant digit of the probability estimate unreliable. In \Cref{fig:rel_err}, we keep the lower integration limits unchanged but use smaller and more dispersed upper integration limits to visualize one example where $10^{4}$ QMC samples are insufficient for keeping the relative errors low. The dense and hierarchical methods are not included because the non-preconditioned methods should have the same error level when using the same QMC sample size. \Cref{fig:rel_err} shows that all methods have a relative error close to or greater than one in $16,384$ dimensions. Nonetheless, the relative error grows more slowly with $n$ for the preconditioned methods, with the iterative block reordering slightly more effective than the non-iterative one. A second observation is that the relative errors of the estimated log-probabilities are on a much smaller magnitude, indicating right-skewness in the distribution of the MVN/MVT probability estimates. Therefore, we may still trust the magnitude of the probability estimates when the relative error is approaching one. The relative errors listed in this paper are the ratios of the Monte Carlo standard errors to the means of the Monte Carlo estimates and for log-probabilities whose errors are not directly available, we estimate the same problem $10$ times to provide replicates of the estimation.

\section{Application to Stochastic Generators}
\label{sec:application}

\subsection{A skew-normal stochastic generator}
One area that benefits from the methods developed in this paper is the likelihood estimation for the statistical models whose probability density function (PDF) involves the MVN/MVT (cumulative) probabilities and in this section, we use a skew-normal stochastic generator to demonstrate this advantage. Stochastic generators model the space-time dependence of the data in the framework of statistics and aim to reproduce the physical process that is usually emulated through a system of partial differential equations. The emulation of the system requires tens of variables and a very fine grid in the spatio-temporal domain, which is extremely time-and-storage demanding \citep{castruccio2016compressing}. For example, the Community Earth System Model (CESM) Large ENSemble project (LENS) required ten million CPU hours and more than four hundred terabytes of storage to emulate one initial condition \citep{jeong2018reducing}. \cite{castruccio2016compressing} found statistical models could become efficient surrogates for reproducing the physical processes in climate science and concluded that extra model flexibilities would facilitate the modeling on a finer scale; see \cite{castruccio2018principles} for a recent account.

The significance of the MVN and MVT methods in this context is an improvement in flexibility by introducing skewness since the majority of statistical models are elliptical. Generally speaking, there are three ways of introducing skewness to an elliptical distribution, all of which involve the CDF of the distribution. The first is through reformulation, which multiplies the elliptical density function by its CDF. The second method introduces skewness via selection, i.e., $(\bX^{\top} , \bY^{\top})^{\top}$ are jointly elliptical and $\bX \mid {\bY>\bmu}$ is a skewed elliptical distribution, where $\bmu$ is the skewness parameter. \cite{arellano2010multivariate} studied the link between PDF reformulation and selection distributions. The third method introduces skewness through constructing a stochastic representation, typically, $\bZ = \bX + |\bY|$, where $\bX$ and $\bY$ are two independent elliptical random vectors. \cite{zhang2010spatial} studied the skew-normal random field based on the third construction and used the Monte Carlo EM \citep{levine2001implementations} algorithm for model selection to avoid the intractable likelihood function.

We use the third method to construct a skew-normal stochastic generator, given its intuitive representation and simplicity for simulation. However, models in this category have intractable PDFs when $\bY$ has a non-trivial covariance structure. To avoid this complexity, we use a diagonal covariance structure for $\bY$ while adding a coefficient matrix to $|\bY|$, whose PDF is tractable based on the properties of $\mathscr{C}$ random vectors developed in \cite{arellano2002definition}. Specifically, a $\mathscr{C}$ random vector can be written as the Hadamard product of two independent random vectors, representing the sign and the magnitude, respectively. When $\bY$ is a $\mathscr{C}$ random vector and $\bX$ is independent from $\bY$, $G(\bX,\bY) \mid \bY > \0$ has the same distribution as $G(\bX,|\bY|)$ for any function $G(\cdot)$. We propose the following stochastic generator that has the flexibility of modeling the correlation between the elliptical component as well as the skewness component:
\begin{equation} 
 \label{equ:sn3mdl}
 \bZ^* = \xi \mathbf{1}_n + \bA \bX + \bB |\bY|.
\end{equation}
Here, $\xi \in \mathbb{R}$ is the location parameter, $\bX$ and $\bY$ are independent standard normal random vectors, and $\bA$ and $\bB$ are parameter matrices. Choosing $G(\bX,\bY)$ to be $\bA\bX+\bB\bY$, the PDF of $\bZ^*$ is tractable and is written explicitly in \Cref{equ:sn_dns} with further parameterizations for $\bA$ and $\bB$.

As a trade-off of the tractable PDF, \Cref{equ:sn3mdl} may have certain drawbacks for modeling skewed distributions. Firstly, its extension to other skewed elliptical distributions, including the skewed MVT distribution, is not straightforward because the sum of independent elliptical distributions may no longer belong to the same distribution family. Secondly, $\bZ^*$ is typically not a random field for most parameterizations of $\bA$ and $\bB$, which makes the inference at new locations difficult. Fortunately, for stochastic generators, the model is usually simulated on a fixed spatial domain without the need for prediction at unknown locations. In this section, we show that the preconditioned TLR QMC method is more suitable for estimating the parameters of \Cref{equ:sn3mdl} than two other state-of-the-art methods and that the fitted skew-normal model is a more realistic stochastic generator for the wind data in Saudi Arabia than the Gaussian random field.

\subsection{Estimation with simulated data}
\label{subsec:est_sim}
In the spirit of parsimony, $\bA$ is assumed to be the lower Cholesky factor from an exponential kernel, $\sigma_1^2 \exp(-\|\bh\|/\beta_1)$ and $\bB$ is assumed as the covariance matrix from another exponential kernel, $\sigma_2^2 \exp(-\|\bh\|/\beta_2)$, both generated from $n$ pre-specified locations on the 2D plane. Hence, there are five parameters in total, $(\xi, \sigma_1, \beta_1, \sigma_2, \beta_2)$. The assumption on $\bA$ makes $\bA \bX$ a MVN distribution, in fact, Gaussian random field, with an exponential covariance structure while $\bB$ is parameterized as a covariance matrix, not a Cholesky factor, for two reasons. Firstly, the row sums of a lower Cholesky factor have great variability in their magnitudes, causing different levels of skewness among the random variables in $\bZ^*$. Secondly, due to the first reason, the likelihood would depend on the ordering of the random variables. When $\bB$ is a covariance matrix, the row sums usually have similar magnitudes and the likelihood function becomes independent from the ordering within $\bZ^*$. The PDF of $\bZ^* = \bz$ can be derived based on the properties of $\mathscr{C}$ random vectors:
\begin{align}
 \label{equ:sn_dns}
 \hspace{-.1cm}2^n \phi_n(\bz - \xi \mathbf{1}_n; \bA\bA^{\top} + \bB\bB^{\top})\Phi_n\{-\bm{\infty} , (\bI_n + \bC^{\top}\bC)^{-1}\bC^{\top}\bA^{-1}(\bz - \xi \mathbf{1}_n) ; (\bI_n + \bC^{\top}\bC)^{-1}\},
\end{align}
with $\bC = \bA^{-1}\bB$. 

\begin{figure}[t]
\begin{center}
\subfloat{
\includegraphics[width = 0.19\linewidth]{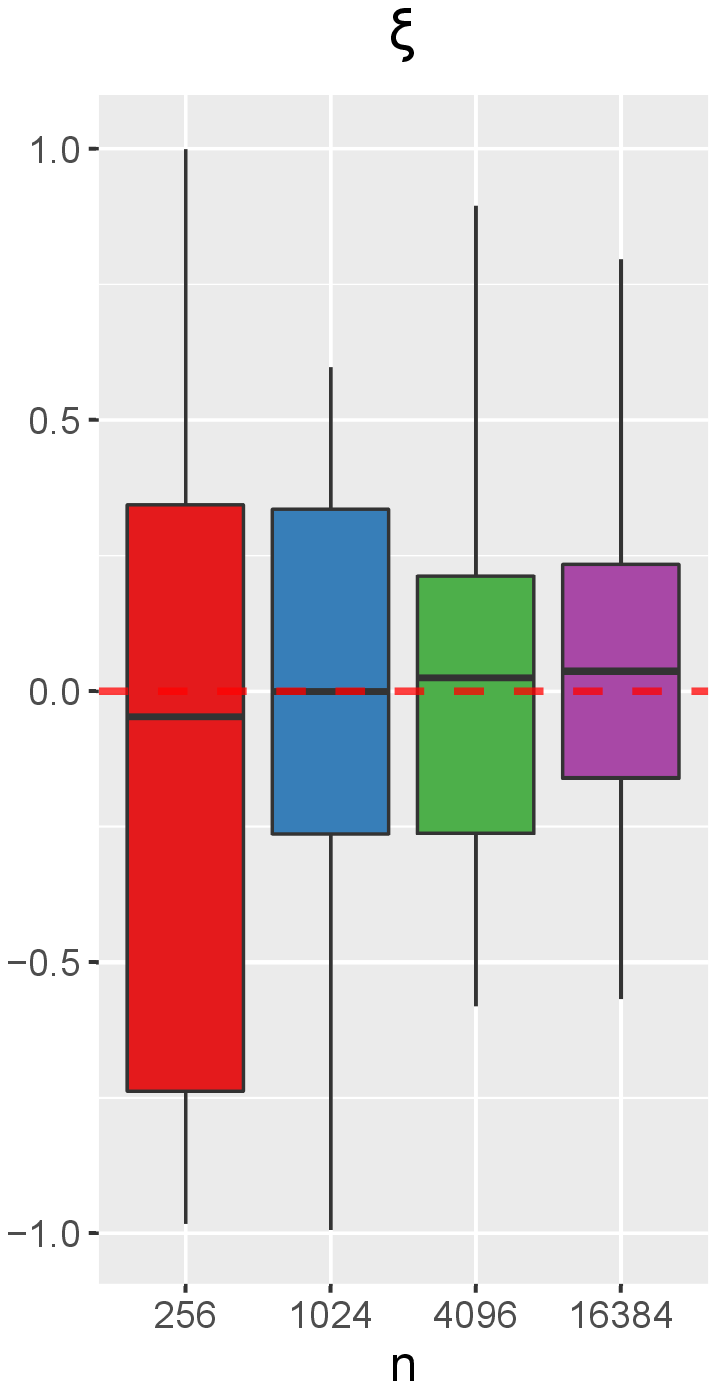}} 
\subfloat{
\includegraphics[width = 0.19\linewidth]{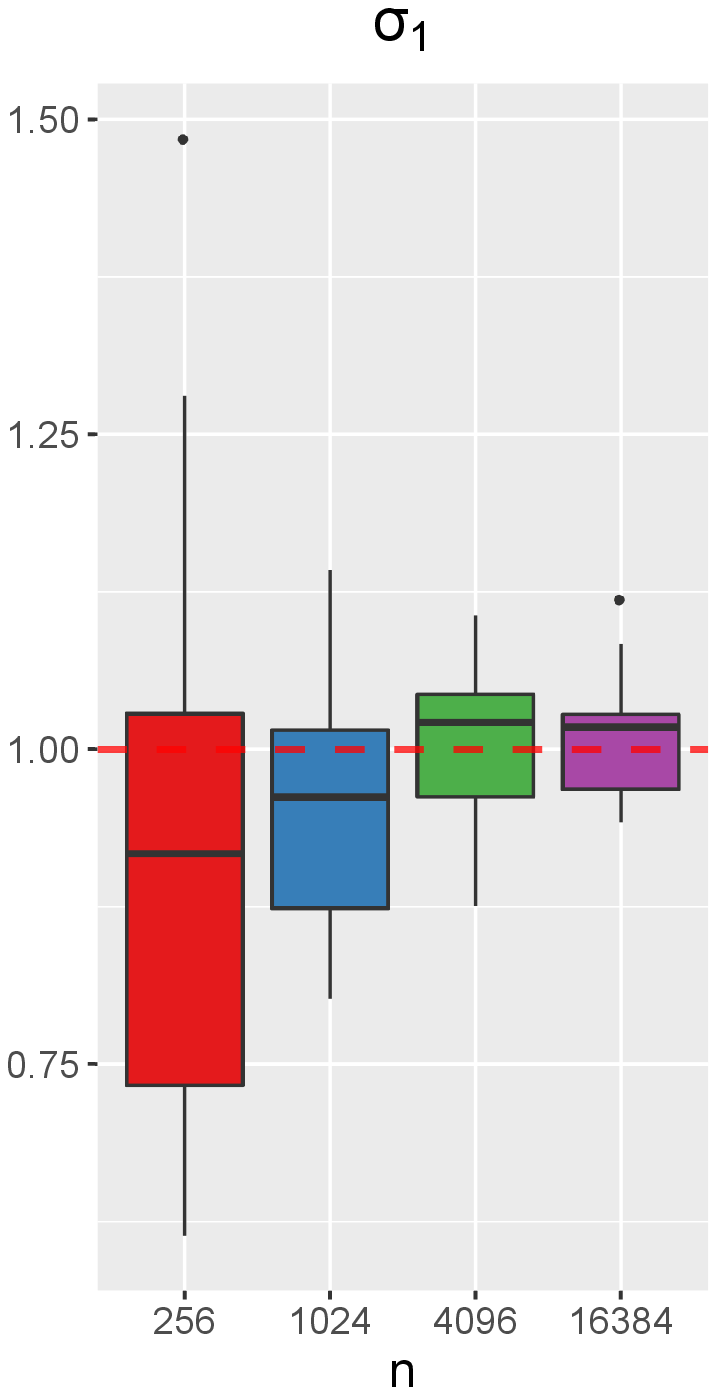}} 
\subfloat{
\includegraphics[width = 0.19\linewidth]{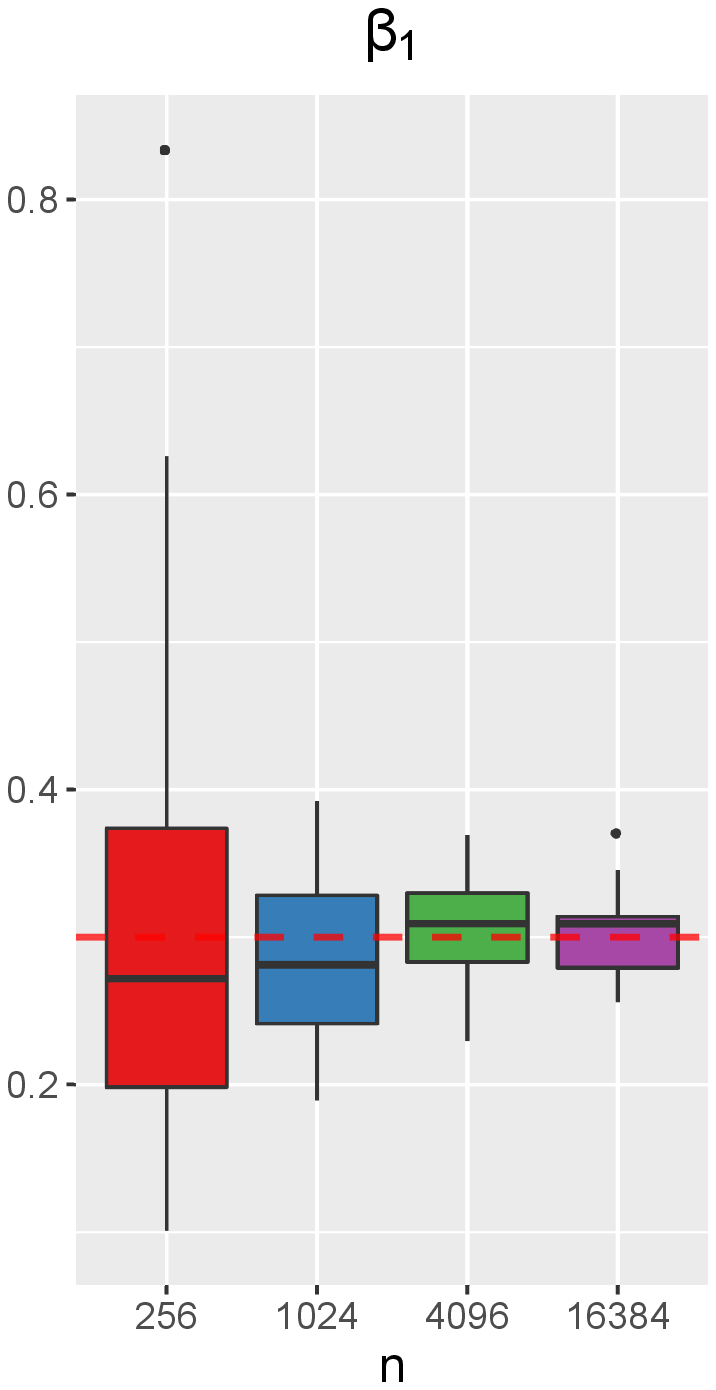}} 
\subfloat{
\includegraphics[width = 0.19\linewidth]{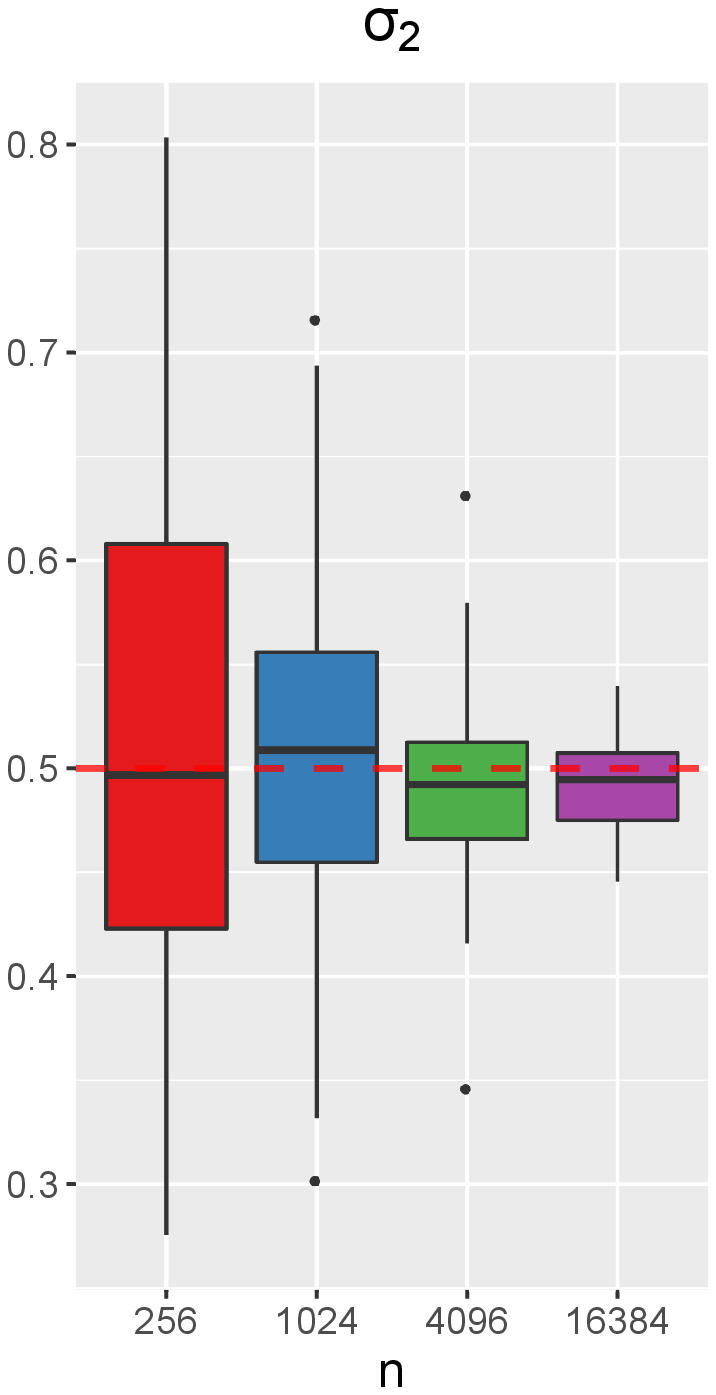}} 
\subfloat{
\includegraphics[width = 0.19\linewidth]{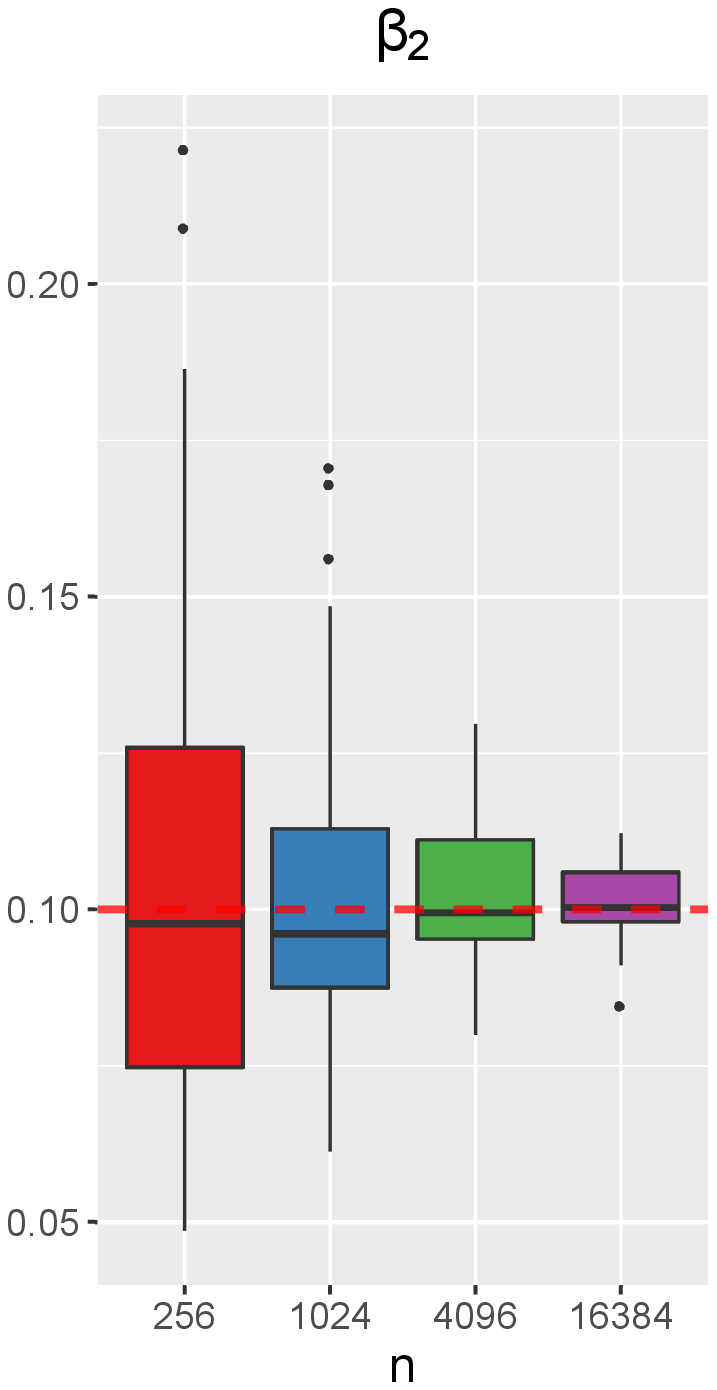}} 
\end{center}
\caption{Boxplots of $30$ estimation results from the maximum likelihood estimators. Each estimation is based on one realization of the $n$-dimensional skew-normal model. The red dashed line marks the true value used for generating random vectors from the skew-normal model.}
\label{fig:est_norm}
\end{figure}

To simulate $\bZ^*$, we first generate $n$ locations on a perturbed grid that expands with $n$. Specifically, for $n = 4^r$ and $r = 4, 5, 6, 7$, a regular grid  in $\mathbb{R}^2$ of dimensions $2^r \times 2^r$ is first generated with the unit distance of $1 / 15$. Then independent disturbances uniformly distributed in $(0, 0.8/15)$ are added to all locations on the grid in both axis orientations to form the perturbed grid, based on which $\bA$ and $\bB$ are constructed with $(\sigma_1 = 1.0, \beta_1 = 0.3)$ and $(\sigma_2 = 1.0, \beta_2 = 0.3)$, respectively. Notice that the ordering of the $n$ locations does not affect the probabilistic distribution of $\bZ^*$. Finally, $\xi$ is assumed zero without loss of generality and $\bZ^*$ is generated based on \Cref{equ:sn3mdl}. For each realization of $\bZ^*$, $\bz \in \mathbb{R}^{n}$, we use the Controlled Random Search (CRS) with local mutation \citep{kaelo2006some}, a global optimization algorithm without gradient usage, to estimate the five parameter values that maximizes \Cref{equ:sn_dns}. In the optimization, the maximum number of iterations is $1{,}000$, the searching ranges for $\{\xi , \sigma_1 , \beta_1 , \sigma_2 , \beta_2\}$ are $\{(-1.0, 1.0), (0.1, 2.0), (0.01, 0.9), (0.0, 1.0), (0.01, 0.3)\}$, respectively and the initial values are the lower bounds of the searching ranges. For each $n$, thirty independent realizations are generated, producing a total of thirty estimation results, which are combined into boxplots in \Cref{fig:est_norm}. Overall, the estimation appears asymptotically unbiased and converges to the true values as the dataset dimension $n$ increases. The outliers indicate that there is sometimes a local maximum, with big $(\sigma_1, \beta_1)$ and small $\sigma_2$, and hence, the fitted skew-normal model is close to a Gaussian random field.

\begin{figure}[b!]
\begin{center}
\subfloat{
\includegraphics[width = 0.19\linewidth]{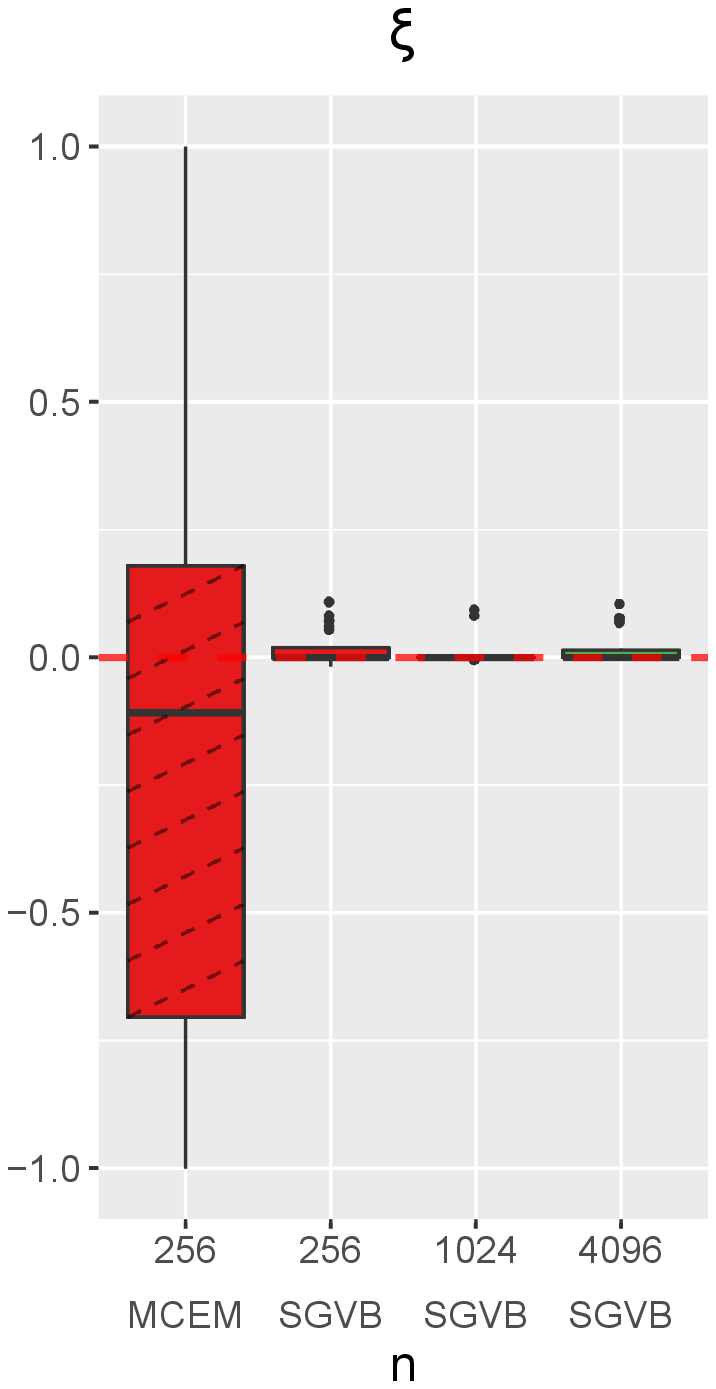}} 
\subfloat{
\includegraphics[width = 0.19\linewidth]{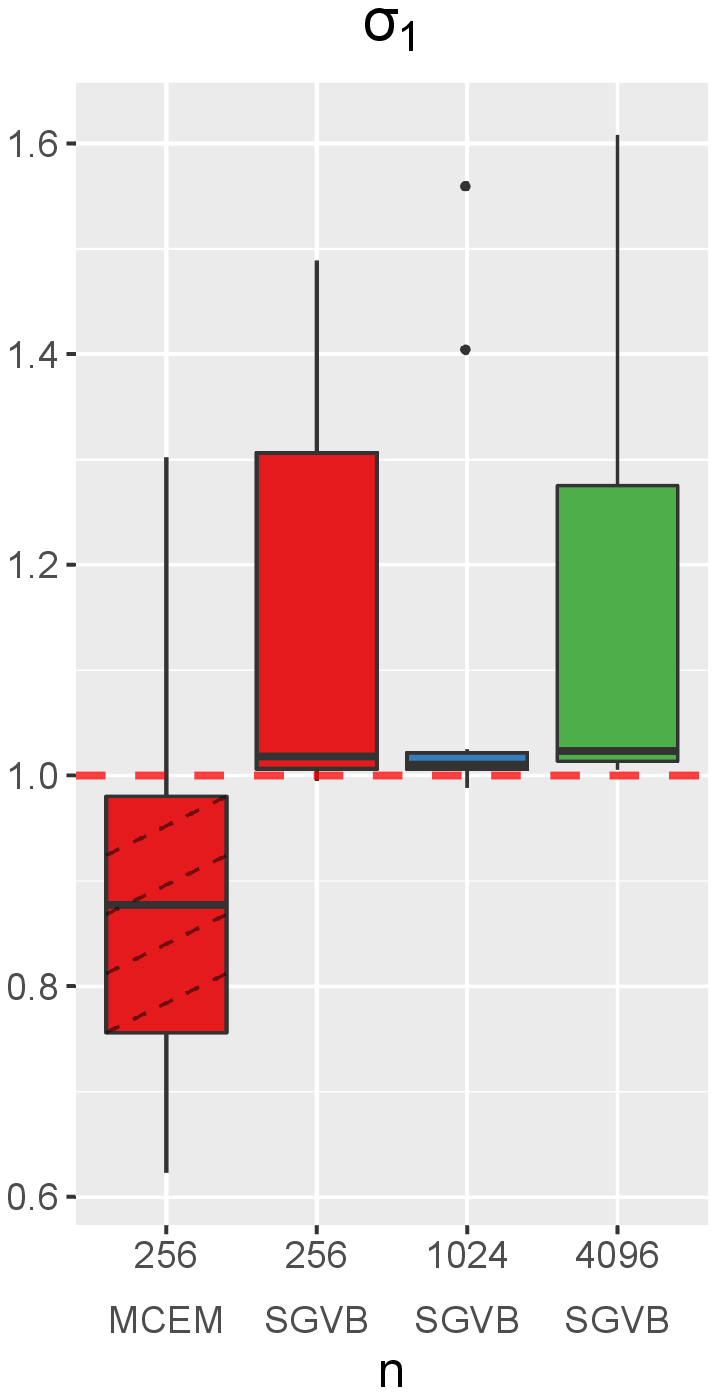}} 
\subfloat{
\includegraphics[width = 0.19\linewidth]{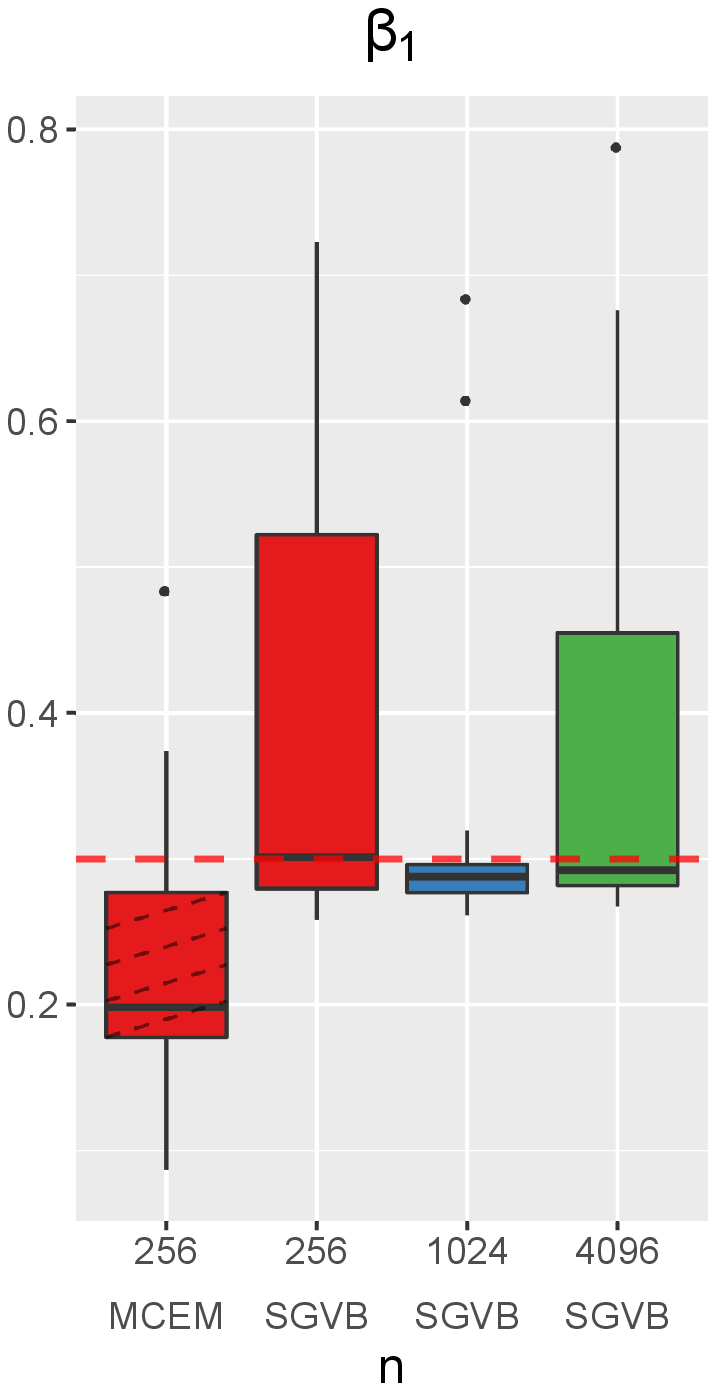}} 
\subfloat{
\includegraphics[width = 0.19\linewidth]{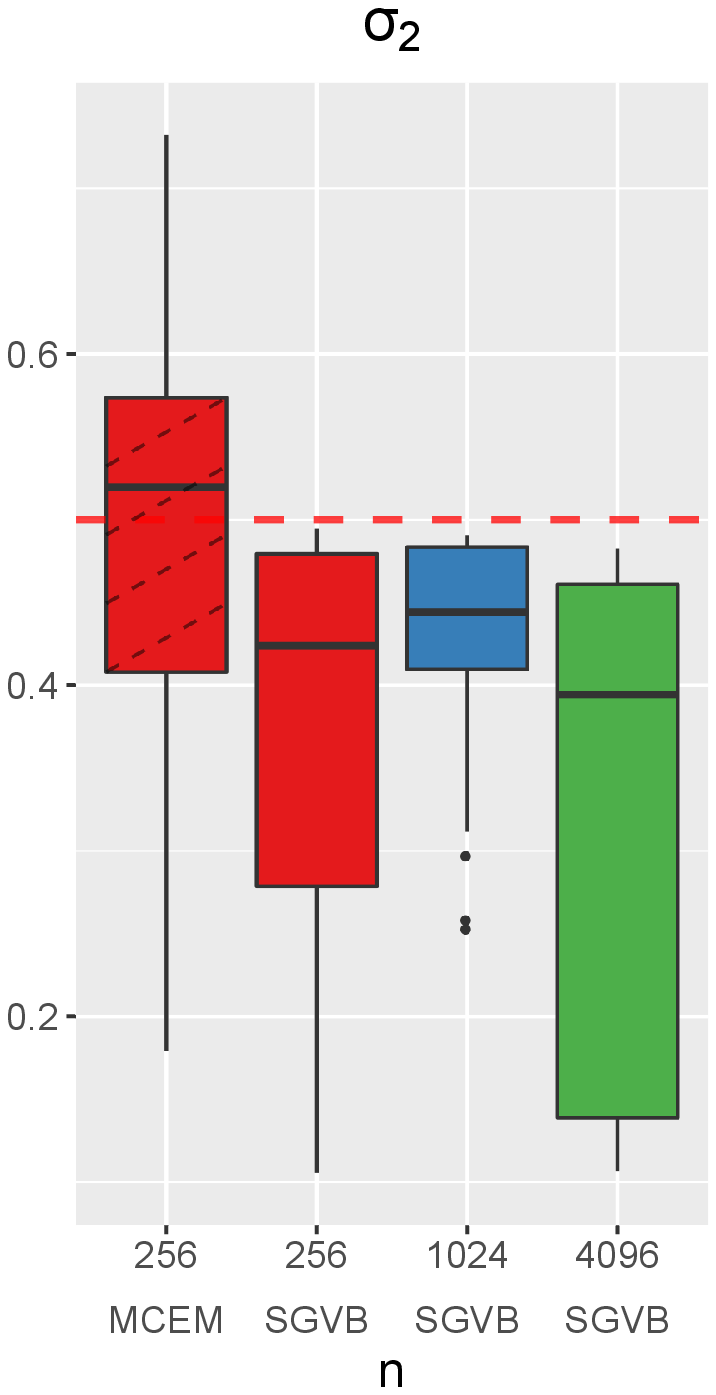}} 
\subfloat{
\includegraphics[width = 0.19\linewidth]{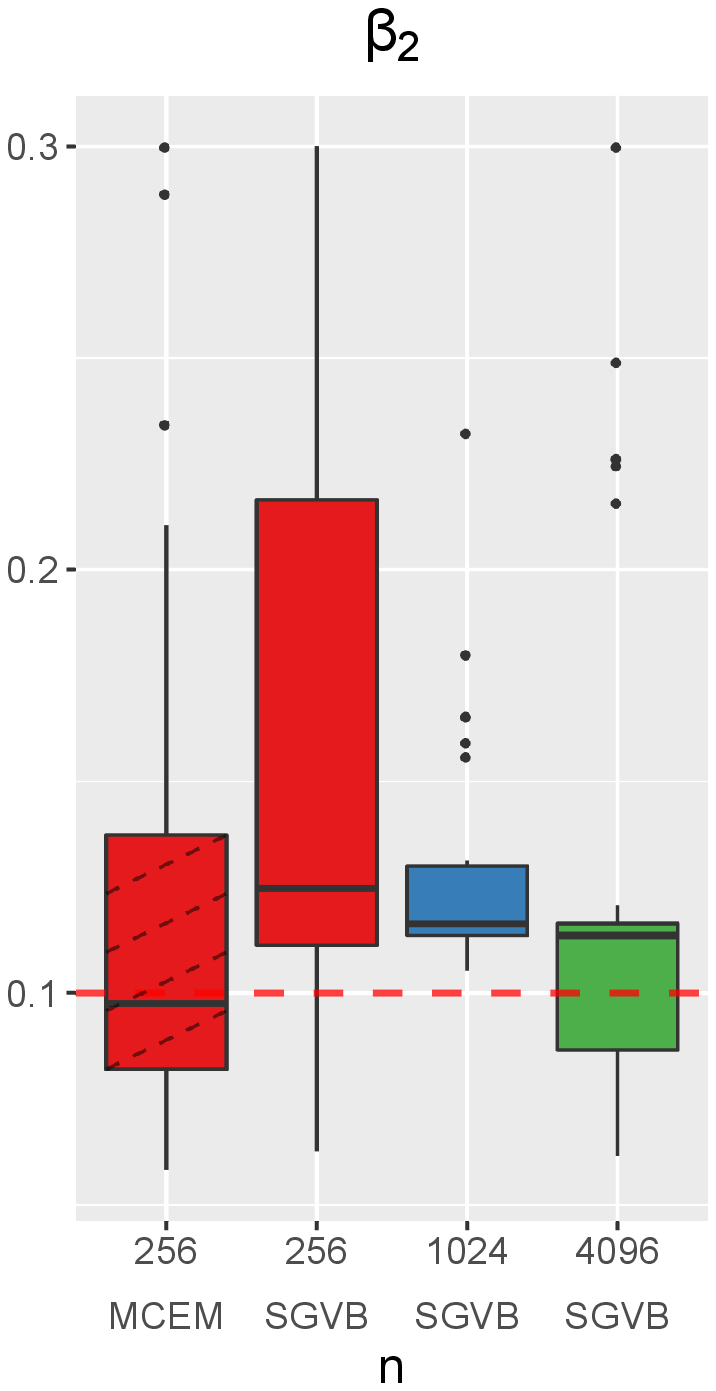}} 
\end{center}
\caption{Boxplots of $30$ estimation results from the Monte Carlo EM (MCEM) estimators (shaded with dashed lines) and the variational Bayes (SGVB) estimators. Each estimation is based on one realization of the $n$-dimensional skew-normal model. The red dashed line marks the true value used for generating random vectors from the skew-normal model.}
\label{fig:sim_study_SGVB}
\end{figure}

As benchmarks, we consider two other state-of-the-art methods frequently used when the likelihood is not tractable or computationally demanding, namely the Monte Carlo EM (MCEM) \citep{levine2001implementations} and the variational Bayes (SGVB) algorithms \citep{kingma2013auto}. In MCEM, sampling techniques, represented by Markov chain Monte Carlo (MCMC) methods, are used in the E-step to approximate the intractable expectation, in our case $E_{\bY \mid \bZ^*, \btheta}[\log f_{\bZ^*, \bY \mid \btheta}(\bz, \by)]$, through sampling from the posterior distribution of the latent random vector, $f_{\bY \mid \bZ^*, \btheta}(\by)$. Here, we use $f(\cdot)$ and $\btheta$ as the general notations for the density function and model parameters, respectively. However, sampling with MCMC can be very expensive and furthermore, this sampling procedure should ideally be repeated in each iteration during optimization, for which only the estimation with $n = 256$ observed locations is performed for our comparison, whose estimated parameters are included in \Cref{fig:sim_study_SGVB}. 
We use the Hamiltonian Monte Carlo method \citep{hoffman2014no} with a burn-in size of $3{,}000$ and a total sample size of $4{,}000$ to sample $f_{\bY \mid \bZ^*, \btheta}(\by)$ and compute $E_{\bY \mid \bZ^*, \btheta}[\log f_{\bZ^*, \bY \mid \btheta}(\bz, \by)]$, which amounts to the logarithm of the MVN PDF. In the M-step, we use the BFGS \citep{nocedal1980updating} algorithm to optimize the five parameters until convergence. We set an upper limit of $10$ iterations between the E-steps and the M-steps, under which the EM algorithm typically cannot reach convergence. To compensate for this, the initial parameter values are set to their true values, under which the MCEM's estimation quality is comparable to that of the maximum likelihood estimators but its computation cost is significantly higher as discussed later in this section.

The variational Bayes algorithm maximizes the variational lower bound as a surrogate of the marginal probability based on an approximate posterior distribution, often variable-wise independent, of the latent random vector $(\bY \mid \bZ^*, \btheta)$; see \cite{kingma2013auto} for more details. Its error from the true marginal probability is the Kullback-Leibler (KL) divergence between the approximate and the true distributions of $(\bY \mid \bZ^*, \btheta)$, changing with $\btheta$ and hence, the optimal parameter values for the marginal probability can be different from those for the variational lower bound. As in \cite{kingma2013auto}, we approximate $f_{\bY \mid \bZ^*, \btheta}(\by)$ with $q_{\bY}(\by) = \prod_{i = 1}^{n}\phi(y_i - \tilde{\xi}_{i}; \tilde{\sigma}_i^2)$ and repeat the simulation study above by maximizing the variational lower bound:
\begin{equation}
  \label{equ:vb_lb}
  -D_{KL}\left\{q_{\bY}(\by) \middle\| f_{\bY \mid \btheta}(\by) \right\} + \int \log\{ f_{\bZ^* \mid \bY, \btheta}(\bz) \} q_{\bY}(\by) \diff \by,
\end{equation}
where $\tilde{\bxi}$ and $\tilde{\bsigma}$ are optimization parameters of length $n$ and $D_{KL}$ is the KL divergence. The KL divergence in \Cref{equ:vb_lb} is tractable while the integration part is evaluated with Monte Carlo samples of $\bY$, which has the same order of complexity as the dense QMC method for MVN probabilities. Given a sample set of $\bY$, \Cref{equ:vb_lb} can be differentiated with respect to its parameters $(\xi, \sigma_1, \beta_1, \sigma_2, \beta_2, \tilde{\bxi}, \tilde{\bsigma})$ explicitly, allowing faster convergence to the local maximum. Similarly to the MCEM algorithm, we also use the gradient-based BFGS algorithm for the optimization with two sets of initial values for $(\xi, \sigma_1, \beta_1, \sigma_2, \beta_2)$, one at their true values and one randomly chosen from their searching ranges, to show that the true values may not be the global minimizer of \Cref{equ:vb_lb}. The initial values for $\tilde{\bxi}$ and $\tilde{\bsigma}$ are $\0$ and $\mathbf{1}$, respectively. In principle, a different starting point can be also added to the MCEM algorithm but this may not be necessary since the experiment for MCEM is confined to $n = 256$ due to high computation costs and its estimation bias is already visible from \Cref{fig:sim_study_SGVB}. Thirty estimation results for each $n$ are summarized in \Cref{fig:sim_study_SGVB}, indicating bias of the variational Bayes estimators as well. Furthermore, the estimation is not necessarily improved after increasing $n$, raising doubts on the consistency of the estimators.

\begin{table}
  \caption{Computation time (seconds) per iteration in the maximum likelihood estimation. Both Monte Carlo EM (MCEM) and Variational Bayes (SGVB) optimize the approximated likelihood in each iteration while MLE-Dense and MLE-TLR optimize the exact likelihood of \Cref{equ:sn_dns} with dense and TLR matrix representations, respectively. For the paired times in MLE-Dense and MLE-TLR, the left is the average time per iteration while the right is the time for computing the MVN probability.}
  \label{tbl:time_MLE_iter}
  \centering
  \smallskip
  \begin{tabular}{ccccc}
    \hline \noalign{\smallskip}
    & $n = 256$ & $n = 1{,}024$ & $n = 4{,}096$ & $n = 16{,}384$ \\
    \hline \noalign{\smallskip}
    MCEM & $4{,}209$s & N.A. & N.A. & N.A.\\
    SGVB & $0.11$s & $2.7$s & $68$s & $1206$s\\
    MLE-Dense & ($0.31$s, $0.28$s) & ($3.0$s, $1.7$s) & ($118$s, $19$s) & ($10{,}784$s, $541$s)\\
    MLE-TLR & ($0.81$s, $0.67$s) & ($6.5$s, $3.9$s) & ($48$s, $17$s) & ($533$s, $104$s)\\
  \end{tabular}
\end{table}

\Cref{tbl:time_MLE_iter} compares the computation costs of the MLE with dense or TLR matrix representations, the MCEM algorithm and the SGVB algorithm. In terms of per optimization iteration, the MLE using the TLR matrix representation (`MLE-TLR') is the fastest but both MLE methods cannot utilize the numerical gradient because the MVN probability estimates are not sufficiently accurate, for which they may need more iterations to converge. Notice that matrix operations, specifically those $O(n^3)$ matrix operations required by \Cref{equ:sn_dns}, account for the major cost of the MLE methods and hence, the `MLE-TLR' method is further amenable to optimized linear algebra libraries for TLR matrices. The MCEM algorithm overall has the highest computation cost due to the MCMC sampling in the E-step, for which its applicability is limited to small dataset dimensions. The SGVB algorithm converges fastest among all and its cost can be further reduced by low-rank matrix representations. However, it may oversimplify the model, causing estimation bias and inconsistency. We proceed with the `MLE-TLR' method for the study of wind data in Saudi Arabia because of its computation feasibility and higher estimation quality.

\subsection{Estimation with wind data from Saudi Arabia}
\label{subsec:est_real}

The dataset we use for modeling is the daily wind speed over a region in the Kingdom of Saudi Arabia on August 5th, 2013, produced by the WRF model \citep{yip2018statistical}, which numerically predicts the weather system based on partial differential equations on the mesoscale and features strong computation capacity to serve meteorological applications \citep{wrfmodel}. The dataset has an underlying geometry with $155$ longitudinal and $122$ latitudinal bands. Specifically, the longitude ranges from $40.034$ to $46.960$ and the latitude ranges from $16.537$ to $21.979$, both with an incremental size of $0.045$. Before fitting the skew-normal model, we subtract the wind speed at each location with its mean over a six-year window (six replicates in total) to increase the homogeneity across the locations. The vectorized demeaned wind speed data is used as the input dataset, $\bZ^*$, for the maximum likelihood estimation. The dataset has a skewness of $-0.45$ and is likely to benefit from the skewness flexibility introduced by the model in \Cref{equ:sn3mdl}. It is worth noting that $\bB |\bY|$ has a negative skewness under our parameterization.

\begin{table}[b!]
 {\footnotesize
 \caption{Parameter specifications and estimations based on the skew-normal (SN) model and the Gaussian random field (GRF)}
 \label{tbl:optinfo_real}
 \begin{center}
 \begin{tabular}{lccccc}
 \hline \noalign{\smallskip}
 & $\xi$ & $\sigma_1$ & $\beta_1$ & $\sigma_2$ & $\beta_2$ \\
 \noalign{\smallskip} \hline \noalign{\medskip}
 Range & $(-2 , 2)$ & $(0.1 , 2.0)$ & $(0.1 , 5.0)$ & $(0.0 , 2.0)$ & $(0.01 , 1.0)$ \\
 \noalign{\smallskip} 
 Initial Value & $0.000$ & $1.000$ & $0.100$ & $1.000$ & $0.010$ \\
 \noalign{\smallskip}
 SN & $-1.211$ & $1.028$ & $4.279$ & $0.419$ & $0.065$ \\
 \noalign{\smallskip}
 GRF & $0.338$ & $1.301$ & $4.526$ & N.A. & N.A. 
 \end{tabular}
 \end{center}
 }
\end{table}

\begin{figure}[t]
\centering
\subfloat[SN]{\includegraphics[trim={0 1.2cm 0 1.9cm}, clip, width = 0.49\linewidth]{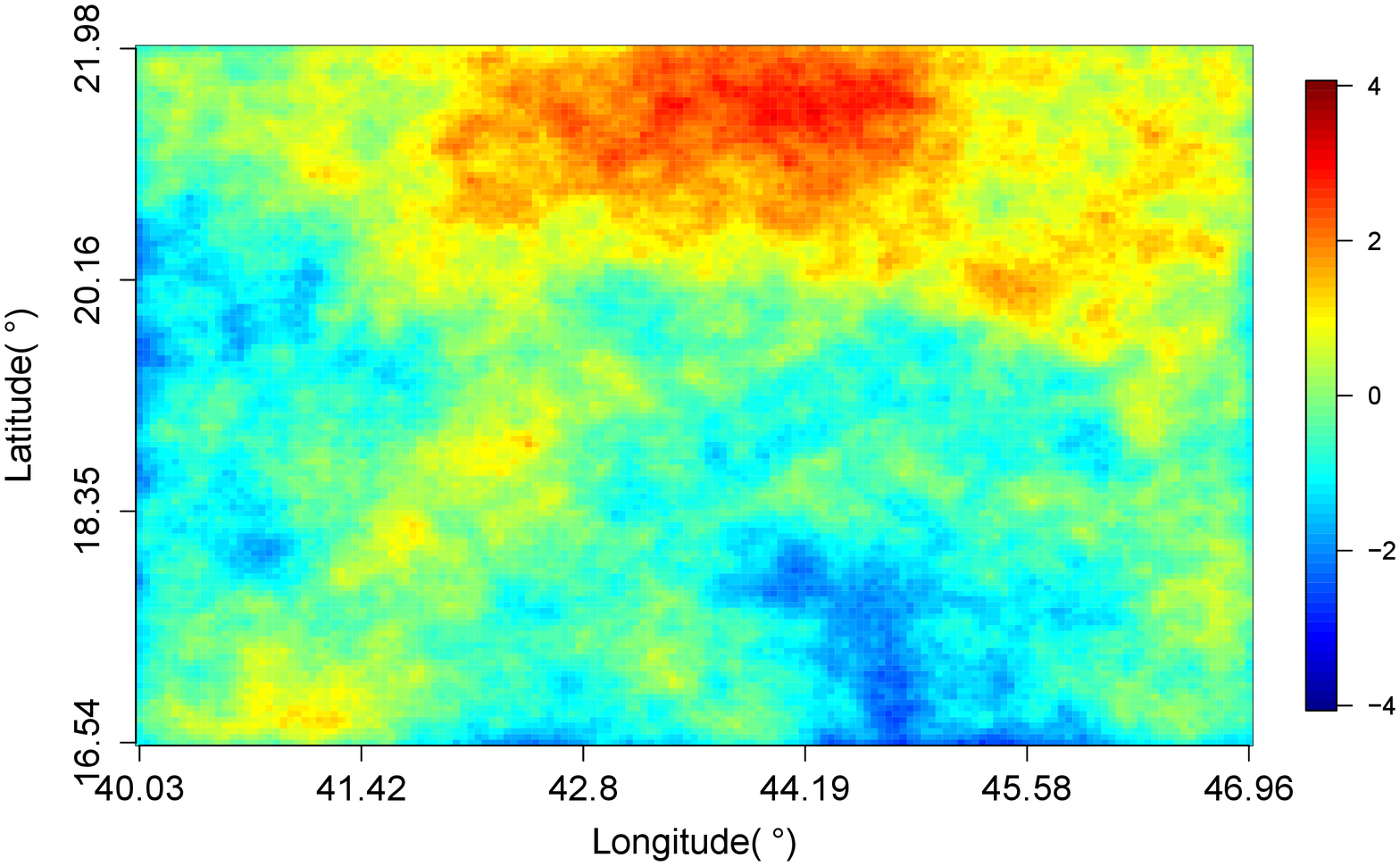}} 
\subfloat[SN]{\includegraphics[trim={0 0.5cm 0 1.7cm}, clip, width = 0.49\linewidth]{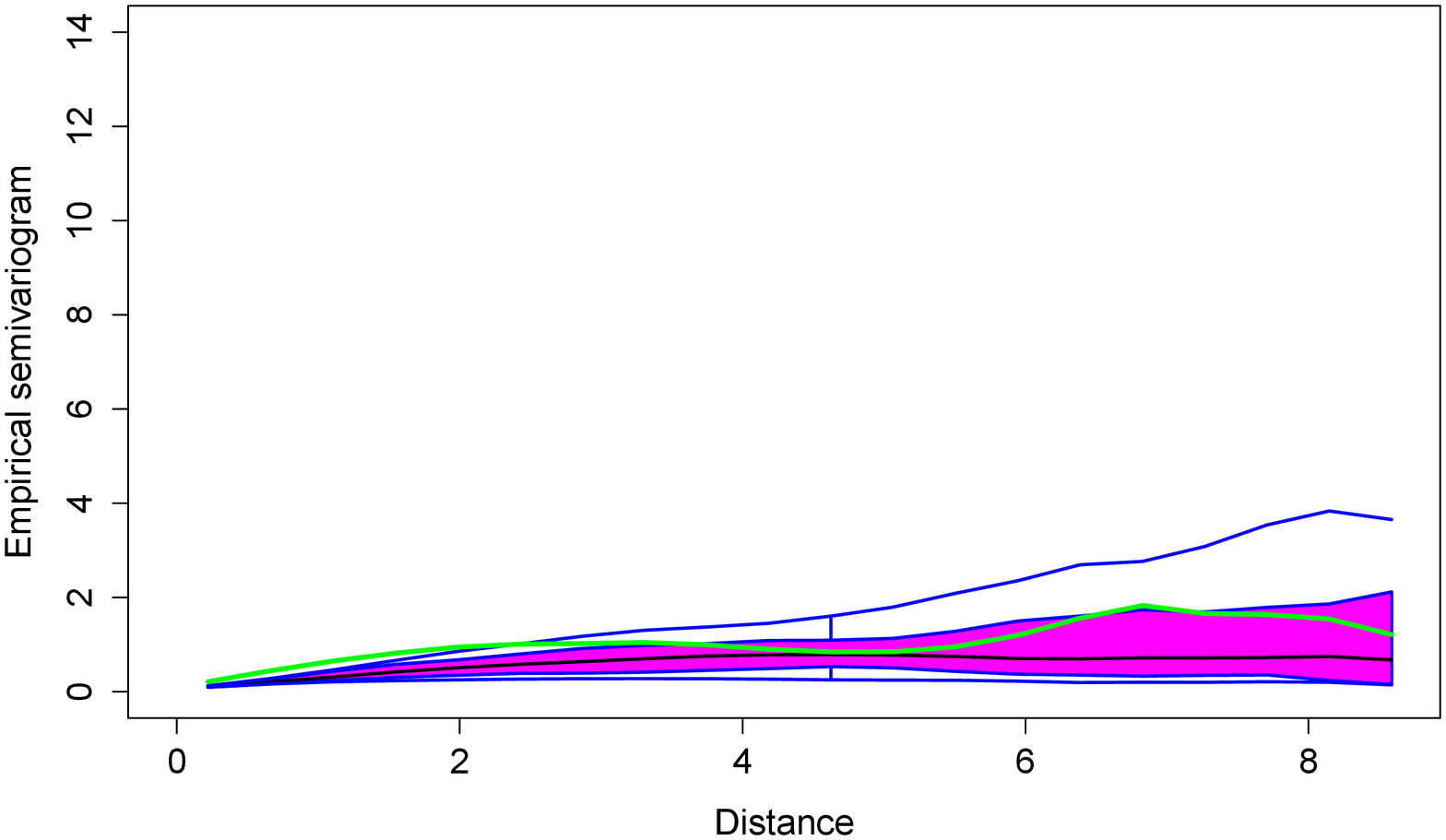}} \\
\subfloat[GRF]{\includegraphics[trim={0 1.2cm 0 1.9cm}, clip, width = 0.49\linewidth]{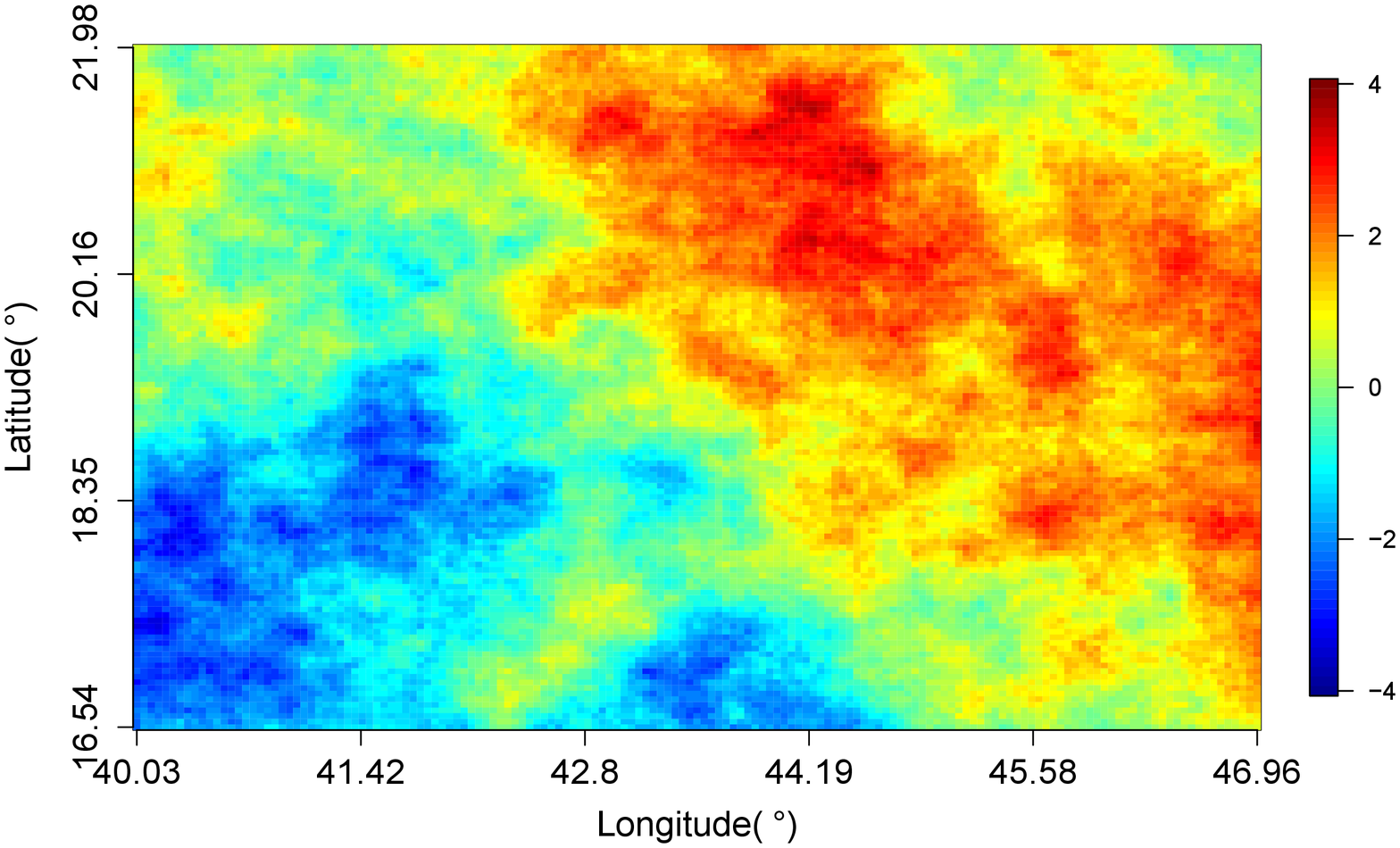}} 
\subfloat[GRF]{\includegraphics[trim={0 0.5cm 0 1.7cm}, clip, width = 0.49\linewidth]{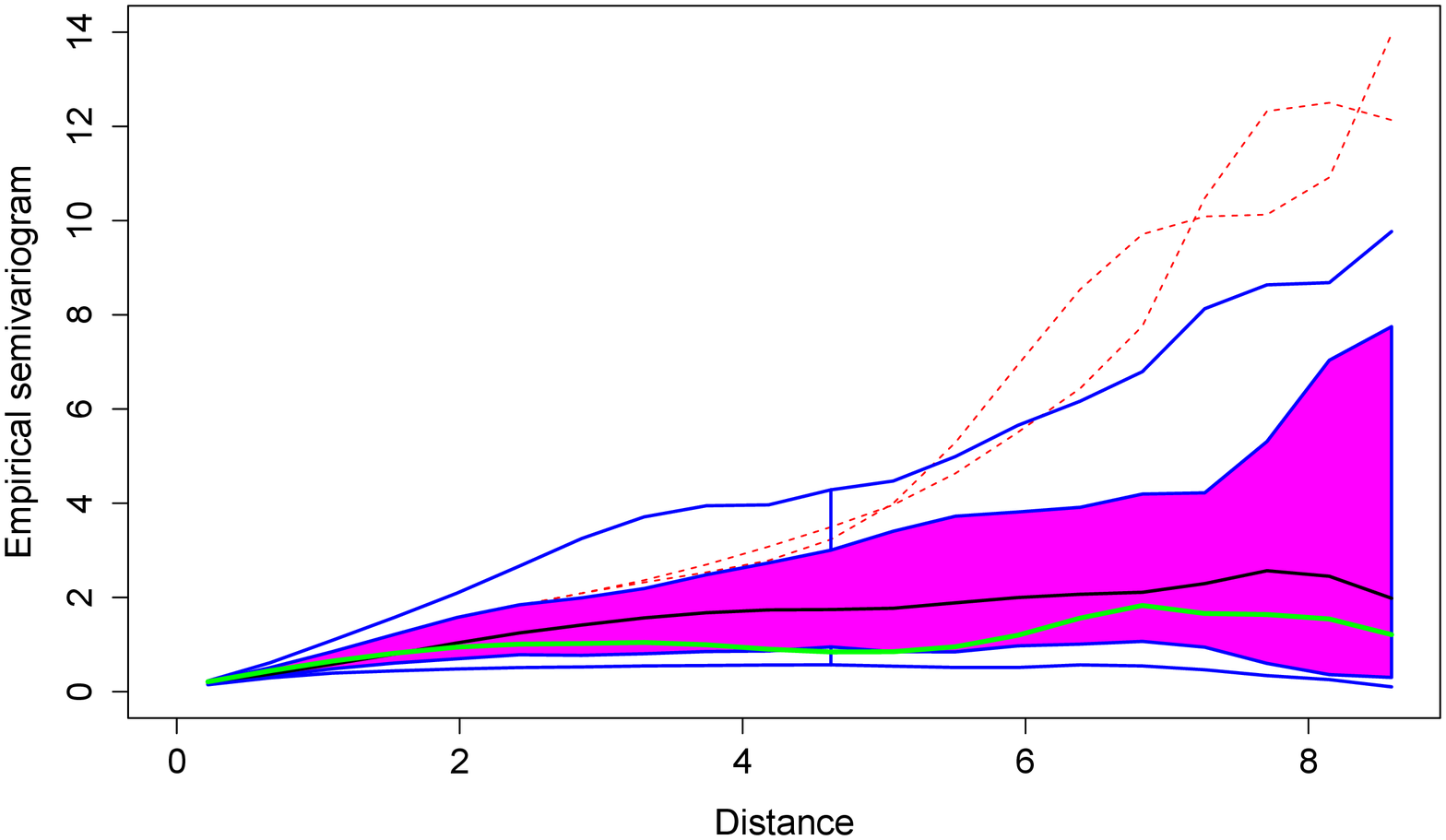}} 
\caption{The heatmap based on one simulation and the functional boxplot of the empirical semivariogram based on $100$ simulations. Top to bottom are the fitted skew-normal model and the Gaussian random field. The green curve denotes the empirical semivariogram based on the wind speed data. The distance is computed as the Euclidean distance in the longitudinal and latitudinal coordinate system.}
\label{fig:variog}
\end{figure}

Same as in \Cref{subsec:est_sim}, the likelihood function is \Cref{equ:sn_dns} and the optimization parameters are $(\xi, \sigma_1, \beta_1, \sigma_2, \beta_2)$. To highlight the contribution of the skewness flexibility, we compare the skew-normal model with the classical Gaussian random field, which is also a special case of the former with $\sigma_2 = 0$. Thus, the Gaussian random field has three optimization parameters $(\sigma_1 , \beta_1 , \xi)$. The initial parameter values, searching ranges and optimized values are summarized in \Cref{tbl:optinfo_real}, where certain lower bounds are above zero to prevent singularity. We first compare the two fitted models with the functional boxplots \citep{sun2011functional} of the empirical semivariogram based on $100$ simulations, which is shown in \Cref{fig:variog}. 
\begin{table}[t]
 {\footnotesize
 \caption{Empirical moments and BIC comparison. SN denotes the skew-normal model and GRF denotes the Gaussian random field. The intervals represent the $5\%$ to $95\%$ quantile intervals based on $100$ simulations.}
 \label{tbl:sn_grf_cmp}
 \begin{center}
 \begin{tabular}{lccccc}
 \hline \noalign{\smallskip}
 & Mean & Variance & Skewness & Kurtosis & BIC\\
 \noalign{\smallskip} \hline \noalign{\medskip}
 Wind data & $0.042$ & $0.932$ & $-0.445$ & $2.873$ & N.A.\\
 \noalign{\medskip}
 SN & $(-1.079 , 1.360)$ & $(0.308 , 1.054)$ & $(-0.644 , 0.449)$ & $(2.274 , 3.595)$ & $22986$ \\
 \noalign{\medskip} 
 GRF & $(-1.644 , 1.911)$ & $(0.612 , 2.594)$ & $(-0.717 , 0.489)$ & $(2.116 , 3.705)$ & $21565$ 
 \end{tabular}
 \end{center}
 }
\end{table}
The skew-normal model has significantly smaller band width than the Gaussian random field while both cover the empirical semivariogram of the original data. Two heatmaps of the fitted models are also shown in \Cref{fig:variog} but their distinction is not as obvious as in the functional boxplots. Next, based on the same $100$ simulations, we compare the quantile intervals of the empirical moments in \Cref{tbl:sn_grf_cmp}, where we also include the BIC values to indicate that the skew-normal model is a better fit. Except for variance, the two models have similar quantile intervals that contain the moments of the wind dataset but since the empirical moments ignore the correlation between the spatial locations, they may not measure the fitting quality in a comprehensive manner.

\section{Conclusion}
\label{sec:conclusion_tlrmvnmvt}

In this paper, we first summarized the SOV methods from \cite{genz1992numerical} and \cite{genz1999numerical} for MVN and MVT probabilities. Two definitions of the MVT probability were compared and one was shown to have better numerical properties than the other. Next, we demonstrated that the TLR structure \citep{weisbecker2013improving, mary2017block, akbudak2017tile} is more aligned with variable reordering than the HODLR structure used in \cite{GKT2016hmvn} as well as the hierarchical structure under the standard admissibility condition, allowing it to benefit from both a reduced cost per sample and an improved convergence rate. Additionally, we introduced an iterative version of the block reordering proposed in \cite{CGKT2018hcmvn} that further improves the convergence rate and produces the TLR Cholesky factor simultaneously. A third contribution is the observation that when the estimation errors are of the same magnitude as the probability estimates, we can still trust the magnitudes of estimates to a certain extent, e.g., for the maximum likelihood estimation of a high-dimensional skew-normal model. 

\section*{Acknowledgements}
The authors thank Prof. Georgiy Stenchikov at KAUST for providing the WRF data and the two anonymous reviewers for valuable comments that improved this manuscript.

\baselineskip=18.5pt
\bibliographystyle{asa} 
\bibliography{mybib}

\begin{thebibliography}{38}
\newcommand{\enquote}[1]{``#1''}
\expandafter\ifx\csname natexlab\endcsname\relax\def\natexlab#1{#1}\fi

\bibitem[{Akbudak et~al.(2017)Akbudak, Ltaief, Mikhalev, and
  Keyes}]{akbudak2017tile}
Akbudak, K., Ltaief, H., Mikhalev, A., and Keyes, D. (2017), \enquote{Tile low
  rank {C}holesky factorization for climate/weather modeling applications on
  manycore architectures,} in \textit{International Supercomputing Conference},
  Springer, pp. 22--40.

\bibitem[{Arellano-Valle et~al.(2002)Arellano-Valle, del Pino, and
  San~Mart{\'i}n}]{arellano2002definition}
Arellano-Valle, R., del Pino, G., and San~Mart{\'i}n, E. (2002),
  \enquote{Definition and probabilistic properties of skew-distributions,}
  \textit{Statistics \& Probability Letters}, 58, 111--121.

\bibitem[{Arellano-Valle et~al.(2006)Arellano-Valle, Branco, and
  Genton}]{arellano2006unified}
Arellano-Valle, R.~B., Branco, M.~D., and Genton, M.~G. (2006), \enquote{A
  unified view on skewed distributions arising from selections,}
  \textit{Canadian Journal of Statistics}, 34, 581--601.

\bibitem[{Arellano-Valle and Genton(2010)}]{arellano2010multivariate}
Arellano-Valle, R.~B. and Genton, M.~G. (2010), \enquote{Multivariate unified
  skew-elliptical distributions,} \textit{Chilean Journal of Statistics}, 1,
  17--33.

\bibitem[{Azzalini and Capitanio(2014)}]{azzalini2013skew}
Azzalini, A. and Capitanio, A. (2014), \textit{The Skew-Normal and Related
  Families}, Cambridge University Press.

\bibitem[{Azzimonti and Ginsbourger(2018)}]{azzimonti2018estimating}
Azzimonti, D. and Ginsbourger, D. (2018), \enquote{Estimating orthant
  probabilities of high-dimensional Gaussian vectors with an application to set
  estimation,} \textit{Journal of Computational and Graphical Statistics}, 27,
  255--267.

\bibitem[{Bolin and Lindgren(2015)}]{bolin2015excursion}
Bolin, D. and Lindgren, F. (2015), \enquote{Excursion and contour uncertainty
  regions for latent Gaussian models,} \textit{Journal of the Royal Statistical
  Society: Series B (Statistical Methodology)}, 77, 85--106.

\bibitem[{B{\"o}rm et~al.(2003)B{\"o}rm, Grasedyck, and
  Hackbusch}]{borm2003introduction}
B{\"o}rm, S., Grasedyck, L., and Hackbusch, W. (2003), \enquote{Introduction to
  hierarchical matrices with applications,} \textit{Engineering Analysis with
  Boundary Elements}, 27, 405--422.

\bibitem[{Botev(2017)}]{botev2017normal}
Botev, Z.~I. (2017), \enquote{The normal law under linear restrictions:
  simulation and estimation via minimax tilting,} \textit{Journal of the Royal
  Statistical Society: Series B (Statistical Methodology)}, 79, 125--148.

\bibitem[{Boukaram et~al.(2019)Boukaram, Turkiyyah, and Keyes}]{boukaram18b}
Boukaram, W., Turkiyyah, G., and Keyes, D. (2019), \enquote{Hierarchical matrix
  operations on {GPU}s: Matrix-vector multiplication and compression,}
  \textit{ACM Transactions on Mathematical Software}, 45, 3:1--3:28.

\bibitem[{Cao et~al.(2019)Cao, Genton, Keyes, and Turkiyyah}]{CGKT2018hcmvn}
Cao, J., Genton, M.~G., Keyes, D.~E., and Turkiyyah, G.~M. (2019),
  \enquote{Hierarchical-block conditioning approximations for high-dimensional
  multivariate normal probabilities,} \textit{Statistics and Computing}, 29,
  585--598.

\bibitem[{Castruccio and Genton(2016)}]{castruccio2016compressing}
Castruccio, S. and Genton, M.~G. (2016), \enquote{Compressing an ensemble with
  statistical models: An algorithm for global 3{D} spatio-temporal
  temperature,} \textit{Technometrics}, 58, 319--328.

\bibitem[{Castruccio and Genton(2018)}]{castruccio2018principles}
--- (2018), \enquote{Principles for statistical inference on big
  spatio-temporal data from climate models,} \textit{Statistics \& Probability
  Letters}, 136, 92--96.

\bibitem[{Durante(2019)}]{durante2019conjugate}
Durante, D. (2019), \enquote{Conjugate {B}ayes for probit regression via
  unified skew-normal distributions,} \textit{Biometrika}, 106, 765--779.

\bibitem[{Genton(2004)}]{genton2004skew}
Genton, M.~G. (2004), \textit{Skew-elliptical Distributions and Their
  Applications: A Journey Beyond Normality}, CRC Press.

\bibitem[{Genton et~al.(2018)Genton, Keyes, and Turkiyyah}]{GKT2016hmvn}
Genton, M.~G., Keyes, D.~E., and Turkiyyah, G. (2018), \enquote{Hierarchical
  decompositions for the computation of high-dimensional multivariate normal
  probabilities,} \textit{Journal of Computational and Graphical Statistics},
  27, 268--277.

\bibitem[{Genz(1992)}]{genz1992numerical}
Genz, A. (1992), \enquote{Numerical computation of multivariate normal
  probabilities,} \textit{Journal of Computational and Graphical Statistics},
  1, 141--149.

\bibitem[{Genz and Bretz(1999)}]{genz1999numerical}
Genz, A. and Bretz, F. (1999), \enquote{Numerical computation of multivariate
  t-probabilities with application to power calculation of multiple contrasts,}
  \textit{Journal of Statistical Computation and Simulation}, 63, 103--117.

\bibitem[{Genz and Bretz(2002)}]{genz2002comparison}
--- (2002), \enquote{Comparison of methods for the computation of multivariate
  t probabilities,} \textit{Journal of Computational and Graphical Statistics},
  11, 950--971.

\bibitem[{Genz and Bretz(2009)}]{genz2009computation}
--- (2009), \textit{Computation of Multivariate Normal and t Probabilities},
  vol. 195, Springer Science \& Business Media.

\bibitem[{Grasedyck et~al.(2008)Grasedyck, Kriemann, and
  Le~Borne}]{grasedyck2008parallel}
Grasedyck, L., Kriemann, R., and Le~Borne, S. (2008), \enquote{Parallel black
  box {$\mathcal{H}$}-LU preconditioning for elliptic boundary value problems,}
  \textit{Computing and Visualization in Science}, 11, 273--291.

\bibitem[{Hackbusch(2015)}]{hackbusch2015hierarchical}
Hackbusch, W. (2015), \textit{Hierarchical Matrices: Algorithms and Analysis},
  vol.~49, Springer.

\bibitem[{Hoffman and Gelman(2014)}]{hoffman2014no}
Hoffman, M.~D. and Gelman, A. (2014), \enquote{The No-U-Turn sampler:
  adaptively setting path lengths in Hamiltonian Monte Carlo.} \textit{Journal
  of Machine Learning Research}, 15, 1593--1623.

\bibitem[{Jeong et~al.(2018)Jeong, Castruccio, Crippa, Genton,
  et~al.}]{jeong2018reducing}
Jeong, J., Castruccio, S., Crippa, P., Genton, M.~G., et~al. (2018),
  \enquote{Reducing storage of global wind ensembles with stochastic
  generators,} \textit{The Annals of Applied Statistics}, 12, 490--509.

\bibitem[{Kaelo and Ali(2006)}]{kaelo2006some}
Kaelo, P. and Ali, M. (2006), \enquote{Some variants of the controlled random
  search algorithm for global optimization,} \textit{Journal of Optimization
  Theory and Applications}, 130, 253--264.

\bibitem[{Kingma and Welling(2013)}]{kingma2013auto}
Kingma, D.~P. and Welling, M. (2013), \enquote{Auto-encoding variational
  bayes,} \textit{arXiv preprint arXiv:1312.6114}.

\bibitem[{Kriemann(2005)}]{kriemann2005parallel}
Kriemann, R. (2005), \enquote{Parallel-matrix arithmetics on shared memory
  systems,} \textit{Computing}, 74, 273--297.

\bibitem[{Levine and Casella(2001)}]{levine2001implementations}
Levine, R.~A. and Casella, G. (2001), \enquote{Implementations of the Monte
  Carlo EM algorithm,} \textit{Journal of Computational and Graphical
  Statistics}, 10, 422--439.

\bibitem[{Mary(2017)}]{mary2017block}
Mary, T. (2017), \enquote{Block low-rank multifrontal solvers: complexity,
  performance, and scalability,} Ph.D. thesis.

\bibitem[{Nocedal(1980)}]{nocedal1980updating}
Nocedal, J. (1980), \enquote{Updating quasi-Newton matrices with limited
  storage,} \textit{Mathematics of Computation}, 35, 773--782.

\bibitem[{Richtmyer(1951)}]{richtmyer1951evaluation}
Richtmyer, R.~D. (1951), \enquote{The evaluation of definite integrals, and a
  quasi-{M}onte-{C}arlo method based on the properties of algebraic numbers,}
  Tech. rep., Los Alamos Scientific Lab.

\bibitem[{Schervish(1984)}]{schervish1984algorithm}
Schervish, M.~J. (1984), \enquote{Algorithm AS 195: Multivariate normal
  probabilities with error bound,} \textit{Journal of the Royal Statistical
  Society. Series C (Applied Statistics)}, 33, 81--94.

\bibitem[{Skamarock et~al.(2008)Skamarock, Klemp, Dudhia, Gill, Barker, Duda,
  Huang, Wang, and Powers}]{wrfmodel}
Skamarock, W.~C., Klemp, J.~B., Dudhia, J., Gill, D.~O., Barker, D.~M., Duda,
  M.~G., Huang, X.-Y., Wang, W., and Powers, J.~G. (2008), \textit{A
  Description of the Advanced Research WRF Version 3}, vol. 113, NCAR.

\bibitem[{Sun and Genton(2011)}]{sun2011functional}
Sun, Y. and Genton, M.~G. (2011), \enquote{Functional boxplots,}
  \textit{Journal of Computational and Graphical Statistics}, 20, 316--334.

\bibitem[{Trinh and Genz(2015)}]{trinh2015bivariate}
Trinh, G. and Genz, A. (2015), \enquote{Bivariate conditioning approximations
  for multivariate normal probabilities,} \textit{Statistics and Computing},
  25, 989--996.

\bibitem[{Weisbecker(2013)}]{weisbecker2013improving}
Weisbecker, C. (2013), \enquote{Improving multifrontal solvers by means of
  algebraic block low-rank representations,} Ph.D. thesis.

\bibitem[{Yip(2018)}]{yip2018statistical}
Yip, C. M.~A. (2018), \enquote{Statistical characteristics and mapping of
  near-surface and elevated wind resources in the Middle East,} Ph.D. thesis,
  King Abdullah University of Science and Technology.

\bibitem[{Zhang and El-Shaarawi(2010)}]{zhang2010spatial}
Zhang, H. and El-Shaarawi, A. (2010), \enquote{On spatial skew-{G}aussian
  processes and applications,} \textit{Environmetrics}, 21, 33--47.

\end{thebibliography}


\end{document}